\documentclass[%
 reprint,
 superscriptaddress,
 amsmath,amssymb,
 aps,
pra,
floatfix,
]{revtex4-2}

\usepackage{graphicx}
\usepackage{verbatim}

\begin{document}


\title{On-demand multimode optical storage in a laser-written on-chip waveguide}

\author{Ming-Xu Su}
\author{Tian-Xiang Zhu}
\author{Chao Liu}
\author{Zong-Quan Zhou}
\email{email:zq\_zhou@ustc.edu.cn}
\author{Chuan-Feng Li}
\email{email:cfli@ustc.edu.cn}
\author{Guang-Can Guo}

\affiliation{CAS Key Laboratory of Quantum Information, University of Science and Technology of China, Hefei 230026, China\\}
\affiliation{CAS Center for Excellence in Quantum Information and Quantum Physics, University of Science and Technology of China, Hefei, 230026, China \\}
\affiliation{Hefei National Laboratory, University of Science and Technology of China, Hefei 230088, China \\}
\date{\today }

\begin{abstract}
Quantum memory is a fundamental building block for large-scale quantum networks. On-demand optical storage with a large bandwidth, a high multimode capacity and an integrated structure simultaneously is crucial for practical application. However, this has not been demonstrated yet. Here, we fabricate an on-chip waveguide in a $\mathrm {^{151}Eu^{3+}:Y_2SiO_5}$ crystal with insertion losses of 0.2 dB, and propose a novel pumping scheme to enable spin-wave atomic frequency comb (AFC) storage with a bandwidth of 11 MHz inside the waveguide. Based on this, we demonstrate the storage of 200 temporal modes using the AFC scheme and conditional on-demand storage of 100 temporal modes using the spin-wave AFC scheme. The interference visibility between the readout light field and the reference light field is $99.0\% \pm 0.6\%$ and $97\% \pm 3\%$ for AFC and spin-wave AFC storage, respectively, indicating the coherent nature of this low-loss, multimode and integrated storage device.

\end{abstract}

\maketitle

\section{Introduction}
\begin{figure*}[htbp]
\centering
\includegraphics[width=0.97\linewidth]{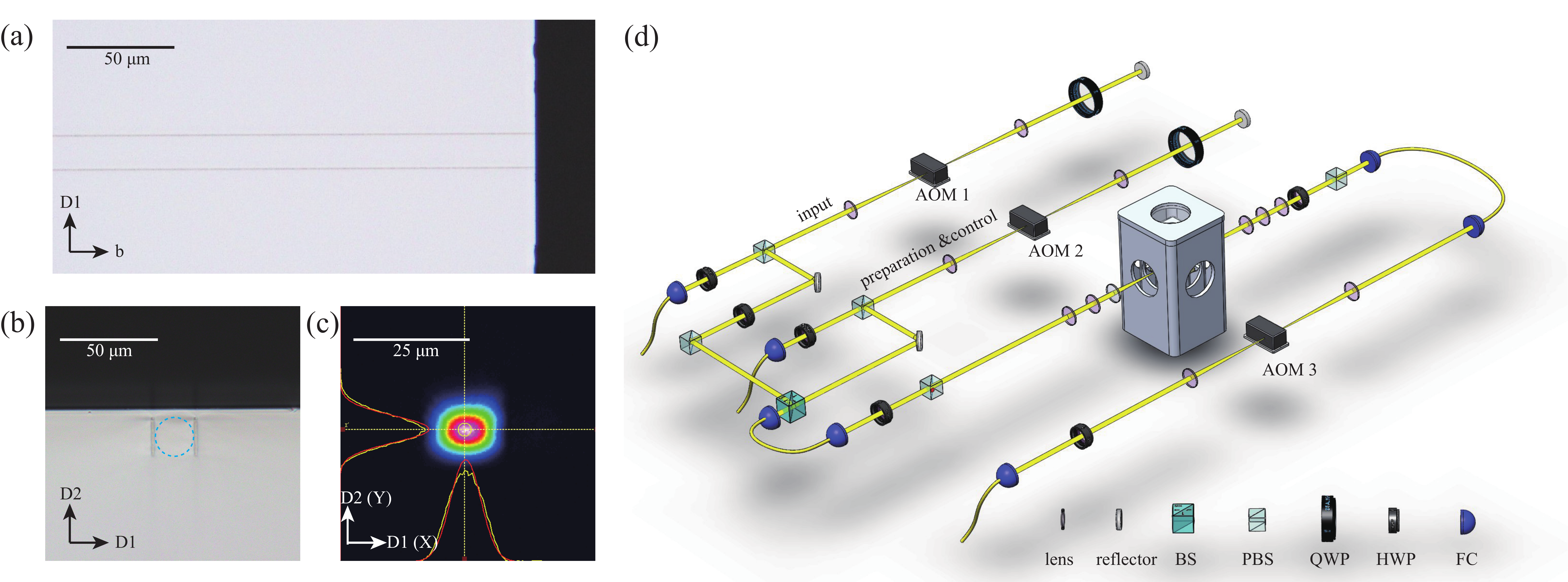}
\caption{ (color online) (a) The top view of the type-II waveguide under an optical microscope. 
(b) The side view of the type-II waveguide under an optical microscope. The blue dashed ellipse indicates the guided regime . 
(c) The intensity profile of guided mode as measured at the exit surface of the waveguide. The full width at half maximum of the guided mode is measured as $\Delta_{\rm{X}}$ = 7.6 $\mu$m and $\Delta_{\rm{Y}}$ = 7.1 $\mu$m. The red lines show the Gaussian fits.
 (d) Experimental setup. The acoustic optical modulator (AOM) 1 and the AOM 2 are in double-pass configurations. The input beam and the preparation and control beam are combined by a beam splitter then coupled into the waveguide. The AOM 3 is in single-pass configuration and serve as a temporal gate. The output beam from AOM 3 was collected by a fibre coupler and sent into a photodetector. Labels at the right corner are: BS = beam splitter, PBS = polarizing beam splitter, QWP = quarter-wave plate, HWP = half-wave plate, FC = fibre coupler.}
    \label{setup}
\end{figure*}
As an interface between light and matter, optical quantum memories (QMs) are of vital importance in quantum information science \cite{gisin2008nature,gisin2011rmp}. Rare-earth-ion (REI) doped crystals are promising candidates for quantum memories because of the long-lived coherence for both the optical transition \cite{Er2009prb} and the hyperfine transition \cite{nature_6hour}, and the large bandwidth \cite{tit2017prl} which enables a 
high multimode capacity.

Multimode storage is a key tool to enhance the data rate of memory-based quantum communication \cite{multimodecomm}. Recently, this property is demonstrated in an elementary quantum repeater segment based on multimode quantum memories \cite{lago2021nature,liu2021experimental}.
The temporal multimode capacity is determined by the time-bandwidth product (storage time$\times$bandwidth) of a memory. For Kramers ions, such as $\mathrm {Nd^{3+}}$, $\mathrm {Er^{3+}}$ and $\mathrm {Yb^{3+}}$, a large storage bandwidth can be obtained due to the large hyperfine splittings. Quantum storage of 100 single photons with a predetermined storage time was demonstrated with $\mathrm{Nd^{3+}}$:$\mathrm{YVO_4}$ \cite{zzq2015nc} and storage of 1060 classical light pulses was demonstrated with $\mathrm{Tm^{3+}}$:$\mathrm{YAG}$ \cite{1060}. However, on-demand spin-wave storage is challenging due to the superhyperfine interactions and fast spin dephasing of these materials. This difficulty was recently overcome by using the zero-field zero first-order Zeema point, on-demand single-mode storage with a bandwidth of 10 MHz is demonstrated with $\mathrm {Yb^{3+}}$:$\mathrm{Y_2SiO_5}$ \cite{Ybspin}. For non-Kramers ions, such as $\mathrm {Eu^{3+}}$ and $\mathrm {Pr^{3+}}$, the nuclear spin states provide long storage time while the bandwidth is typically limited to 10 MHz \cite{spin2010,Eumuti2015Q,Euspin2015Q,Prspin2015Q}. $\mathrm {^{151}Eu^{3+}}$:$\mathrm{Y_2SiO_5}$ is a unique material that has enabled optical storage for 1 hour \cite{ma2021onehour} and the longest spin coherence time among all matter systems \cite{nature_6hour}. On-demand optical storage of 50 temporal modes has been demonstrated in this material with a storage bandwidth of 5 MHz \cite{EU100-50}.

Integrated optical QMs are required in order to connect with other integrated devices and to build the quantum network with integrated architectures. Many techniques have been used to fabricate integrated optical memories based on REI doped crystals, such as femtosecond laser micromachining (FLM) \cite{lminter,reid2018optica,frqmulti2019,lc2020optica,ztxpra,lcprl}, focused-ion-beam milling \cite{fara2017sci,fara2019prap,kindem2020control}, Ti indiffusion in $\mathrm {LiNbO_3}$ waveguide \cite{tit2017prl,gisin2007prl,tit2007prl,tit2011nature,tit2014prl,tit2019prap} and optical fibres \cite{fiberNC,fiberprl}. FLM has the advantages of low damage to the substrate and three-dimensional fabrication capability \cite{CFwaveguide}. Multimode quantum storage of 15 frequency modes and 9 temporal modes with a predetermined storage time has been demonstrated in a laser-written type-I waveguide based on $\mathrm {Pr^{3+}:Y_2SiO_5}$ \cite{frqmulti2019}. Recently, our group demonstrated the on-demand qubit storage in a type-IV on-chip waveguide fabricated on $\mathrm {Eu^{3+}}$:$\mathrm{Y_2SiO_5}$ \cite{lcprl}.

Here, we fabricate a low-loss type-II waveguide close to the surface of a $\mathrm {^{151}Eu^{3+}}$:$\mathrm{Y_2SiO_5}$ crystal. We demonstrate on-demand optical storage with a bandwidth of 11 MHz and a multimode capacity of 100. The reliability of this device is confirmed by the high interference visibility ($99.0\% \pm 0.6\%$ for AFC and $97\% \pm 3\%$ for spin-wave AFC) between the readout light field and the reference light field. 

\section{Fabrication of the waveguide}

The substrate is a 0.1\% doped $\mathrm {^{151}Eu^{3+}}$:$\mathrm{Y_2SiO_5}$ crystal, with a dimension of $15 \times 4 \times 3 $ mm $(b \times D_1 \times D_2)$ and an isotopic enrichment of 99\% for $\mathrm {^{151}Eu^{3+}}$. The configuration of the FLM waveguide is defined as four types according to the relative position of the tracks and reflective-index changes introduced by the femtosecond laser \cite{CFwaveguide}. Here we fabricated a type-II waveguide very close to the substrate's surface, which enables the interface with other on-chip integrated devices. The type-II waveguide is fabricated by a FLM system from WOPhotonics (Altechna R$\&$D Ltd, Lithuania). The femtosecond laser with a wavelength of 1030 nm is injected on the crystal along the $D_2$ axis through a microscope objective (50 $\times$, 0.65 NA). In order to minimize transmission loss, the laser parameters are optimized as follows: pulse duration of 210 fs, energy-per-pulse of 64 nJ, repetition rate of 201.9 kHz, polarization along the \emph{b} axis, and the sample moves along the \emph{b} axis at a speed of 1 mm/s. Note that the energy-per-pulse of 64 nJ is much smaller than that used in our previous work \cite{lc2020optica} which results in smoother sidewalls and lower losses. The resulted damage track is 7 $\mu$m along the femtosecond laser propagation direction ($D_2$ axis). To match the mode of input Gaussian beam, a waveguide with a height and a width of 21 $\mu$m is fabricated. Each edge is formed by three tracks with a depth increment of 7 $\mu$m. After this, a single-mode waveguide for 580-nm laser is fabricated near the surface of the crystal (see Fig. \ref{setup}). The femtosecond laser causes a decrease of refractive-index in the irradiated regions, so the 580-nm laser polarized along the $D_1$ axis is bounded in the waveguide region, while the laser polarized along the $D_2$ axis can not be confined. After the fabrication, the crystal is annealed in air at 800°C for 4 hours and cooled to room temperature in 8 hours. This annealing process reduces the insertion loss (calculated as the ratio of transmission of the waveguide regime and that of the bulk regime of the same crystal, see more details in Appendix A) of the waveguide from 0.4 dB to 0.2 dB. The full width at half maximum (FWHM) of the guided mode is measured as $\Delta_{\rm{X}}$ = 7.6 $\mu$m and $\Delta_{\rm{Y}}$ = 7.1 $\mu$m (X represents 
the direction along the $D_1$ axis, Y represents the direction along the $D_2$ axis)[Fig. \ref{setup} (c)]. After the fabrication process, we measured the coherence time $T_2$ using two-pulse photon echo experiments. The optical coherence lifetime $T_2$ of the excited state is measured as 241 $\mu$s for waveguide regime with a peak power of the input pulse of 0.13 mW. For comparison, the optical coherence lifetime $T_2$ in bulk regime is 267 $\mu$s with a peak power of the input pulse of 0.98 mw. These results indicate that the waveguide fabrication process does not affect the coherence property of the sample.

\section{EXPERIMENTAL SETUP}
A frequency-doubled semiconductor laser (TA-SHG, Toptica) serves as the laser source, which is stabilized to a high-finesse Fabry-Perot cavity at the frequency of 516.849 THz with a linewidth of sub-kHz. The $\mathrm {^{151}Eu^{3+}}$:$\mathrm{Y_2SiO_5}$ crystal is cooled to approximately 3.1 K in a closed-cycle cryostat (Montana Instrument) and mounted on a three-axis translation stage to facilitate the waveguide coupling. The laser is modulated by the acoustic optical modulators (AOMs) driven by an arbitrary waveform generator. As shown in Fig. \ref{setup} (d), the signal beam is generated by AOM 1, which can control the phase and the amplitude of the input pulse. The AOM 2 is employed to generate the preparation and control beam. Those two beams are combined by a 90:10 beam splitter and coupled into the waveguide by three lenses. By adjusting those three lenses, beam diameter is adjusted to match the mode of the waveguide (see Appendix A). A fibre coupler collects the output beam into a single-mode fibre. The output pulses is finally detected by a photodetector after passing a temporal gate based on the AOM 3. The overall transmission efficiency from the front of the cryostat to the single-mode fibre is 56\%. Since the main loss is due to the reflections of the uncoated sample surfaces and other optical elements, higher efficiencies can be obtained with better antireflection coatings.

\section{RESULT}
\begin{figure*}[htbp]
\centering
 \includegraphics[width=0.95\linewidth]{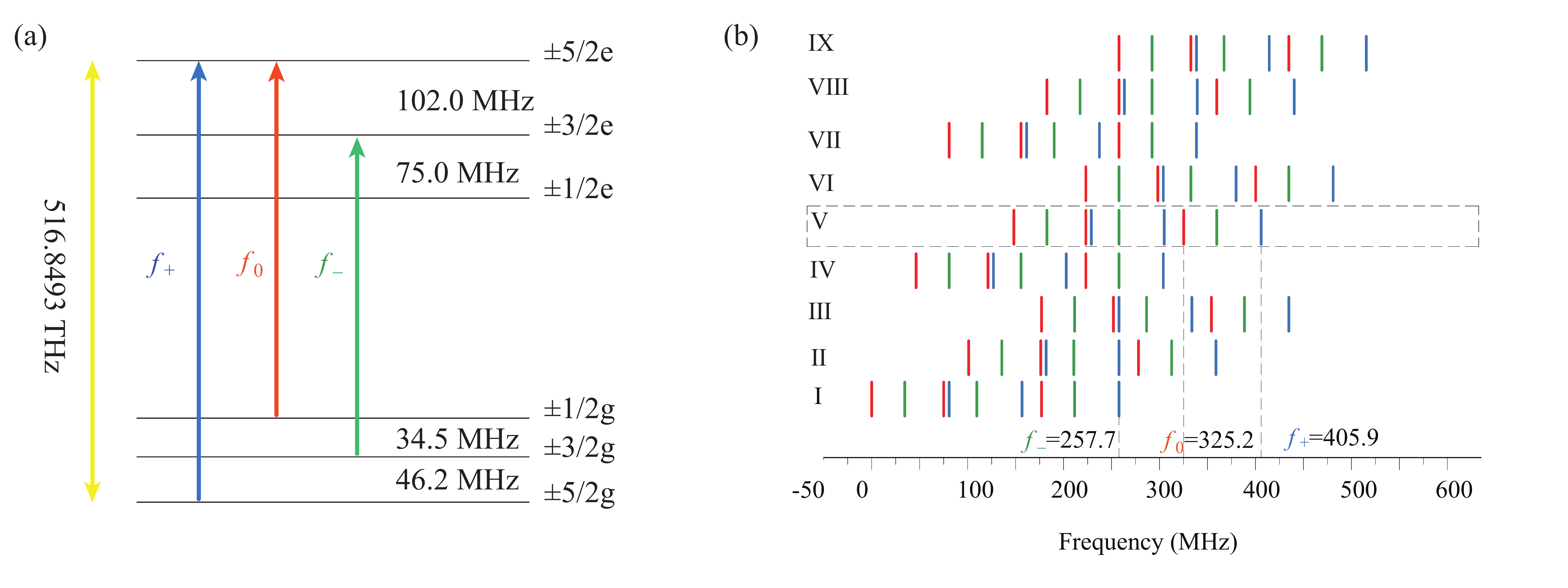}
\caption{ (color online) (a) The energy level diagram of the $^7F_0$ $\rightarrow $ $^5D_0$ transition of $\mathrm {^{151}Eu^{3+}}$ in $\mathrm {Y_2SiO_5}$ crystal. The input pulses with the center frequency of $f_0$ are resonant with ${\left | \pm {1/2} \right \rangle}_g\rightarrow{\left | \pm {5/2} \right \rangle}_e$, and the control pulses with the center frequency of $f_+$ are resonant with ${\left | \pm {5/2} \right \rangle}_g\rightarrow{\left | \pm {5/2} \right \rangle}_e$. Pulses with the center frequency of $f_-$ are necessary in the spectral preparation process. (b) Stick diagram of all 81 transitions for ions of class I-IX. The red sticks represent the transitions from ${\left | \pm {1/2} \right \rangle}_g$. The green sticks represent the transitions from ${\left | \pm {3/2} \right \rangle}_g$. The blue sticks represent the transitions from ${\left | \pm {5/2} \right \rangle}_g$. The relative frequency of each transition can be calculated from the energy level diagram (see Appendix B). Ions of class V are selected for the experiment. The maximum bandwidth of 11.7 MHz is limited to the gap (11.7 MHz) between the third green stick of class-VIII ions and the third blue stick of class-V ions.}
\label{levelpump}
\end{figure*}
\subsection{The preparation of an 11-MHz $\Lambda$-system in ${^{151}Eu^{3+}}$:$\mathrm{Y_2SiO_5}$}
The atomic frequency comb (AFC) is an established protocol for spin-wave quantum storage in REI doped solids \cite{Eumuti2015Q,Euspin2015Q,Prspin2015Q}. Spectral-hole burning is employed to prepare a comb-like absorption structure (with a periodicity of $\Delta$) in the inhomogeneously broadened optical transition \cite{Euspec,spin2010}. Input pulse absorbed by this structure will dephase and rephase, leading to a photon-echo-like emission at time 1/$\Delta$ \cite{gisin2009pra}. In order to achieve on-demand and long-lived storage, the spin-wave AFC scheme can be implemented, which requires applying a pair of control pulses before the two-level AFC echo appears. The first control pulse is applied to map the excitation into the spin state. After the time $T_s$, the second control pulse brings it back and continue the AFC rephasing. The total storage time T = 1/$\Delta$ + $T_s$.

 The multimode capacity in the time domain of the AFC scheme is proportional to the number of comb teeth \cite{gisin2009pra}, which depends on the total bandwidth and the periodicity $\Delta$. While for spin-wave AFC memory, the storage bandwidth will be further limited by hyperfine splittings, depending the chosen $\Lambda$-system. So far, this bandwidth is limited to 5 MHz in $\mathrm {^{151}Eu^{3+}}$:$\mathrm{Y_2SiO_5}$ \cite{EU100-50,Eumuti2015Q,Euspin2015Q,lc2020optica,ztxpra}. In order to overcome this limitation, we choose a different $\Lambda$-system from previous works, which has a larger bandwidth of 11 MHz but a lower transition probability \cite{Euspec} for the control pulses. However, the manipulation of such weak transition can be achived inside a waveguide due to the strong confinement of light in the small interaction area.

\begin{figure*}[htbp]
\centering
\includegraphics[width=\linewidth]{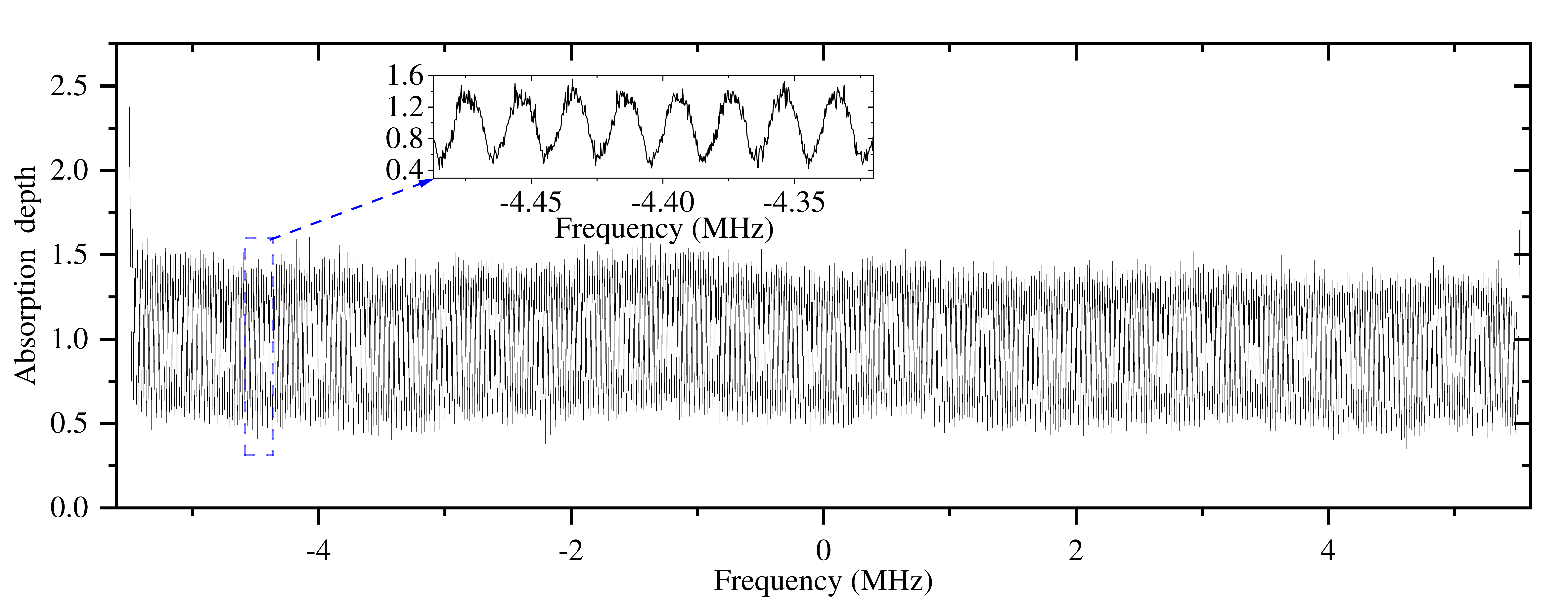}
\caption{The measured AFC structure with a bandwidth of 11 MHz and periodicity of 20 kHz. The frequency is represented by the detuning to $f_0$. There are 550 teeth in total. The peak absorption and background absorption is approximately 1.05 and 0.44 respectively. The finesse is approximately 2.3. The inset shows the details of the AFC structure.}
\label{AFCstructure}
\end{figure*}

The inhomogeneous broadening of $^7F_0$ $\rightarrow $ $^5D_0$ transition is approximately 3 GHz, which is much larger than the hyperfine splitting (Fig. \ref{levelpump}). Nine classes of ions will interact with the input laser [Fig. \ref{levelpump} (b)]. To implement spin-wave AFC scheme, optical pumping is required to isolate a $\Lambda$-system involving a single class of ions \cite{Eumuti2015Q}. To maximize the useful bandwidth of the $\Lambda$-system, we design a novel strategy for optical pumping. Setting the frequency of transition ${\left | \pm {1/2} \right \rangle}_g\rightarrow{\left | \pm {1/2} \right \rangle}_e$ as zero, those transitions with the frequency of $f_-$ = 257.7 MHz, $f_0$ = 325.2 MHz and $f_+$ = 405.9 MHz are shown in Fig. \ref{levelpump} (b). The first step is to simultaneously apply two 20-MHz sweeping pulses with center frequencies of $f_0$ and $f_+$ to pump all populations into the ground state ${\left | \pm {3/2} \right \rangle}_g$ and create two transparent pits in ${\left | \pm {1/2} \right \rangle}_g$ and ${\left | \pm {5/2} \right \rangle}_g$. The second step is applying 11-MHz sweeping pulses with the center frequency of $f_-$, to pump back ions into the pit in ${\left | \pm {1/2} \right \rangle}_g$. Meanwhile, 11-MHz sweeping pulses with the center frequency of $f_+$ are applied to clean the ${\left | \pm {5/2} \right \rangle}_g$. After those preparation sequences, ions of class V are selected. The bandwidth is limited to 11.7 MHz because the transition ${\left | \pm {3/2} \right \rangle}_g\rightarrow{\left | \pm {5/2} \right \rangle}_e$ of class-VIII ions will mix with the ${\left | \pm {5/2} \right \rangle}_g$ of class-V ions [Fig. \ref{levelpump} (b)]. In the first two steps, each pulse has a duration of 1 ms and repeats for 10 times. The third step is to create the comb on ${\left | \pm {1/2} \right \rangle}_g$ $\rightarrow$ ${\left | \pm {5/2} \right \rangle}_e$ by using the parallel preparation sequences \cite{EU100-50}. The preparation pulse has a duration time of 5 ms and repeats for 76 times. An 11-MHz sweeping pulse with the center frequency of $f_+$ is applied after every two preparation pulses to keep the ${\left | \pm {5/2} \right \rangle}_g$ empty. Finally, the AFC with a bandwidth of 11 MHz and a periodicity of 20 kHz is created with the measured structure presented in Fig. \ref{AFCstructure}.

\subsection{The multimode AFC storage }
\begin{figure}[htbp]
\centering
\includegraphics[width=0.95\linewidth]{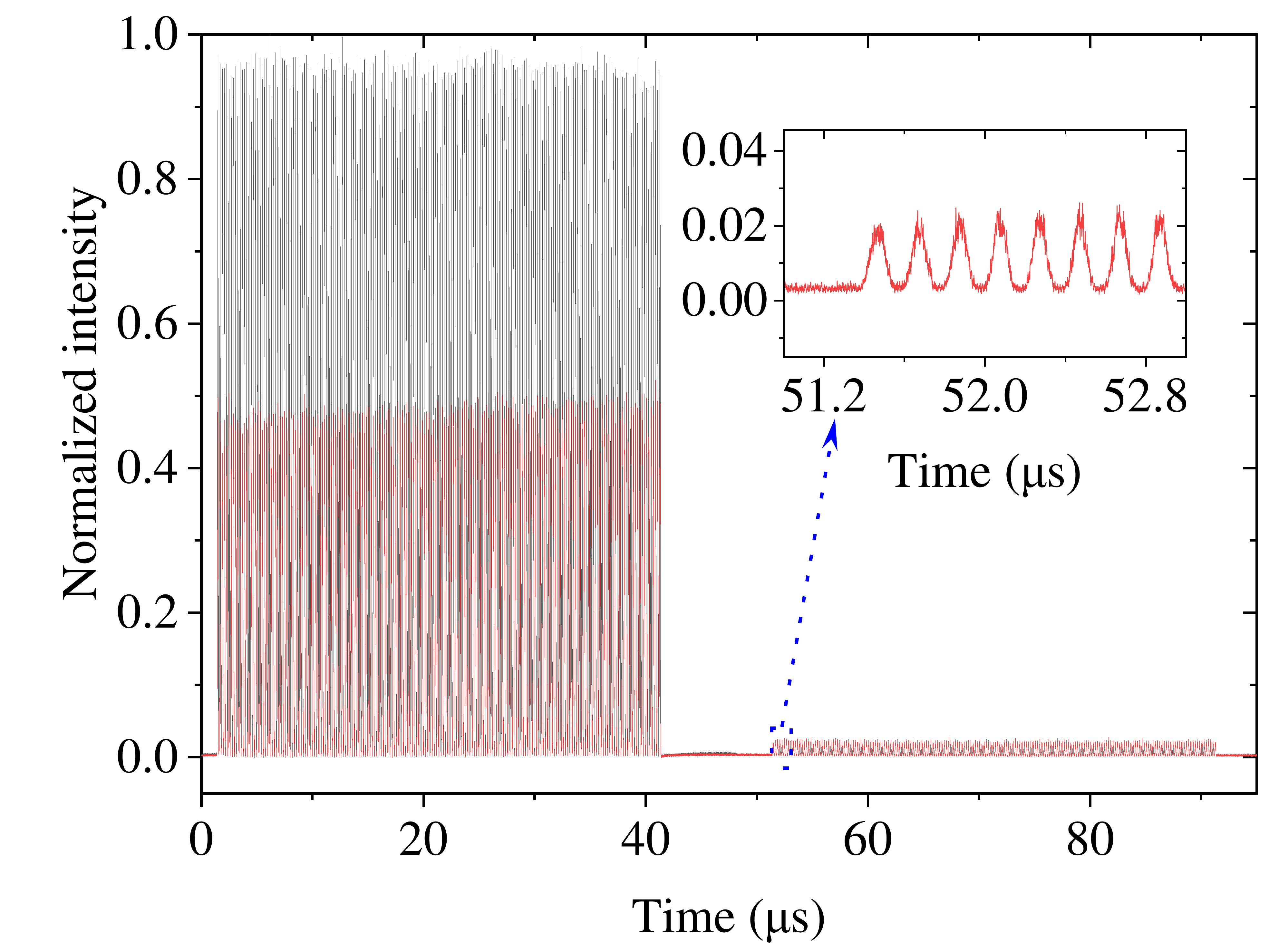}
\caption{ (color online) AFC storage of 200 temporal modes. Each input pulse (black line) has a mode size of 200 ns, matching the AFC bandwidth. The red line shows the trace for AFC storage with a storage time of 50 $\mu$s. The average storage efficiency over 200 modes is $2.5\% \pm 0.2\%$. The inset shows the details about the first 8 modes after storage. }
\label{200}
\end{figure}
 Here, we perform a multimode AFC storage of 200 temporal modes with low crosstalks. As shown in Fig. \ref{200}, 200 temporal modes are injected and each input mode has a mode size of 200 ns and a full width at half maximum (FWHM) of approximately 70 ns, which matches the bandwidth of 11 MHz. After 1/$\Delta$ = 50 $\mu$s, AFC echoes appear. Each AFC echo has a FWHM of approximately 73 ns and a mode size of approximately 200 ns, which indicates that nearly all of the input mode is stored. 
The AFC efficiency is defined as the ratio between the intensity of the AFC echo and the input pulse. The retrieved AFC echoes have an average efficiency of $2.5\% \pm 0.2\%$. This value agrees with the theoretical storage efficiency of 2.3\% estimated from the AFC structure by using the formula $\eta = e^{-\tilde{d}}\tilde{d}^{2}e^{-7/F^{2}}e^{-d_{0}}$ \cite{gisin2008nature}. Here, $\tilde{d} = d/F$, d is the peak absorption depth, $d_{0}$ is the background absorption depth, F is the finesse of the AFC, and these parameters are provided in Fig. \ref{AFCstructure}. Assuming d = 1.49 and a background absorption $d_{0}$ = 0, the highest theoretical efficiency of 11.3 \% can be predicted with a square-shaped comb peak with a finesse of 2.35 \cite{gisin2009pra,PhysRevA.81.033803}. The reduced efficiency in our experiment is due to the imperfect pumping caused by some technical issues (finite laser linewidth, vibration of the cryostat), which can be especially significant for AFC with large bandwidth and long storage times.

\begin{figure}[htbp]
\centering
\includegraphics[width=0.95\linewidth]{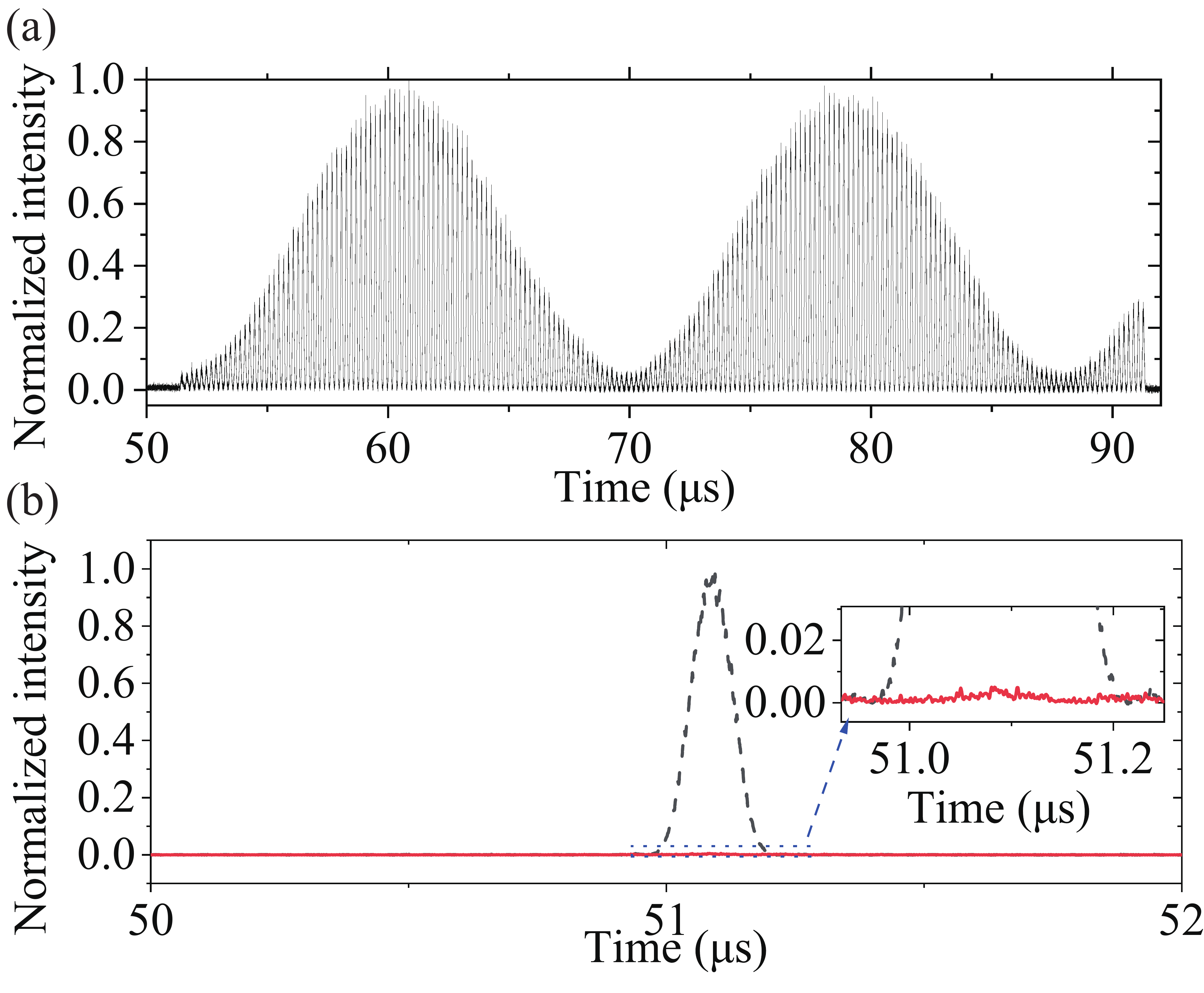}
\caption{ (color online) (a) Interference between 200 echoes and 200 reference pulses in AFC storage. The phase of each reference pulse increases by four degrees. (b) The traces of the readout echo in the situations of the constructive interference (black dashed line) and the destructive interference (red solid line) which are averaged by ten measurements. The interference visibility is $99.0\% \pm 0.6\%$. The inset shows the details about the destructive interference.}
\label{AFC}
\end{figure}

To verify the coherent nature of this highly multimode memory, 200 synchronized reference pulses are injected into the waveguide along the same path as the input signal pulses. The phase of each pulse is increased by 4 degrees one by one while the phase of all input pulses are the same. The interference of the echo and reference pulses form a trigonometric-function-like pattern [Fig. \ref{AFC} (a)], showing that the coherence is well preserved over all modes. Because it is hard to tune every reference pulses to be exactly the same as every echo, there are some remaining signals in the location with a destructive interference.

For preciser measurement, the interference visibility is further measured by operating with only a single mode to precisely tune the mode to overlap between the echo and the reference pulse [Fig. \ref{AFC} (b)]. The reference pulse is tuned to the same shape and same time as the echo but different in phase. The interference visibility is calculated from $V = \frac{max-min}{max+min}$, where $max$ and $min$ corresponding to the constructive interference intensity and the destructive interference intensity. The visibility is measured to be $V = 99.0\% \pm 0.6\%$.

\subsection{The multimode spin-wave AFC storage}
\begin{figure}[htbp]
\centering
 \includegraphics[width=0.95\linewidth]{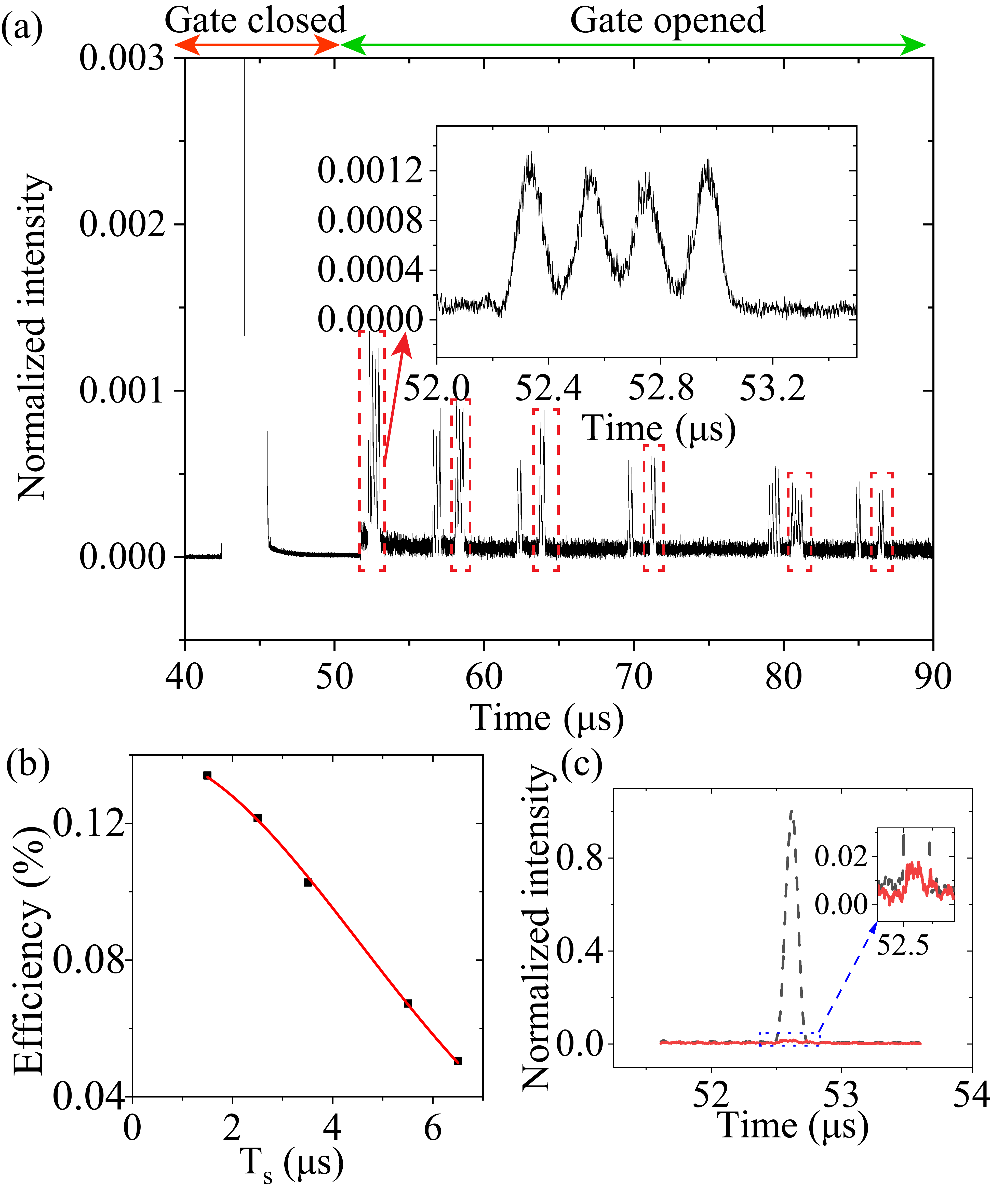}
\caption{ (color online) (a) Spin-wave AFC storage of 100 modes for 51.5 $\mu$s. The AOM gate is opened when the spin-wave AFC echoes emit. The peaks in red dashed line are the spin-wave AFC echoes. Other peaks are the remaining part of the AFC echoes. The inset shows the details about the first four modes. (b) The spin-wave AFC efficiency as a function of spin-wave storage time ($T_s$). The red solid line is a Gaussian function fit with the fitted inhomogeneous spin linewidth of 58.8$\pm$0.5 kHz. (c) The traces of spin-wave echo in the situation of the constructive interference (black dashed line) and the destructive interference (red solid line) averaged by ten measurements. The visibility is $97\% \pm 3\%$. The inset shows the details about the destructive interference.}
\label{spin}
\end{figure}
Finally, we demonstrate the temporal multimode memory based on the spin-wave AFC scheme. Compared to the AFC scheme, spin-wave AFC scheme could enable on-demand and long-lived storage. After the AFC is prepared, we inject occupied and empty temporal modes into the waveguide. Then, two control pulses are applied before the AFC echo appears. The control pulses have a power of 140 mW, a chirp bandwidth of 11 MHz and a duration time of 1.5 $\mu$s. We demonstrate the conditional on-demand storage of 100 temporal modes with a storage time of 51.5 $\mu$s (Fig. \ref{spin}). The spin-wave AFC echo has a FWHM of approximately 96 ns. The average efficiency of the first four modes is $0.13\% \pm 0.01\%$, which is calculated as the ratio between the intensity of the spin-wave AFC echo and the input pulse. Assuming the spin broadening as a Gaussian disruption, the efficiency can be written as the formula $\eta\left(T_{\mathrm{s}}\right)_{SW}=\eta(0)_{SW} \times \exp \left[\frac{-\left(\gamma_{\mathrm{inh}} T_{\mathrm{s}}\right)^{2}}{2 \ln2 / \pi^{2}}\right]$, where $\gamma_{\mathrm{inh}}$ represents the inhomogeneous spin linewidth \cite{timoney2012atomic,gundougan2013coherent}. By changing the time interval $T_s$ of the two control pulses and recording the amplitude of the output echo, $\gamma_{\mathrm{inh}}$ is determined as 58.8$\pm$0.5 kHz [see Fig. \ref{spin} (b)]. A previous work \cite{21kPhysRevB.89.184305} measured the inhomogeneous broadening is 21 kHz for transition ${\left | \pm {1/2} \right \rangle}_g\leftrightarrow{\left | \pm {3/2} \right \rangle}_g$ in a bulk crystal. The spin Hamiltonian can be written as $H=D\left[I_{z}^{2}-\frac{I(I+1)}{3}\right]$, where D represents a parameter corresponds to the distribution of crystal-field \cite{21kPhysRevB.89.184305,prPhysRevB.84.104417}. The transition energies can be calculated as 2D and 6D for $I_{z}={\left | \pm {1/2} \right \rangle}_g\leftrightarrow{\left | \pm {3/2} \right \rangle}_g$ and $I_{z}={\left | \pm {1/2} \right \rangle}_g\leftrightarrow{\left | \pm {5/2} \right \rangle}_g$, respectively. Since the inhomogeneous broadening is proportional to the transition energy, we can estimate the inhomogeneous broadening of transition ${\left | \pm {1/2} \right \rangle}_g\leftrightarrow{\left | \pm {5/2} \right \rangle}_g$ is 63 kHz, three times that of transition ${\left | \pm {1/2} \right \rangle}_g\leftrightarrow{\left | \pm {3/2} \right \rangle}_g$. This value agrees with measured $\gamma_{\mathrm{inh}}$ = 58.8$\pm$0.5 kHz in our experiment, which indicates that the waveguide fabrication process does not affect the spin inhomogeneous broadening. Using the formula ${T_{2}^{*}}=\frac{\sqrt{2 \ln 2}}{\pi\gamma_{\mathrm{inh}}}$, $T_2^*$ is deduced as $6.4 \pm 0.1$ $\mu$s. 
The transfer efficiency $\eta_{\mathrm{T}}$ of each control pulse can be calculated from the formula $\eta_{\mathrm{SW}}=\eta_{\mathrm{AFC}} \eta_{\mathrm{T}}^{2}$ \cite{2010prlc}. The efficiency without spin dephasing $\eta_{\mathrm{SW}}=$ 0.14\% can be deduced from Fig. \ref{spin}(b). The AFC efficiency $\eta_{\mathrm{AFC}}$ = 2.5\%. Therefore the spin transfer efficiency $\eta_{\mathrm{T}}$ = 23.7\%. Note that increasing the duration of control pulse can improve the transfer efficiency, but due to the limitation of $T_2^*$, the storage efficiency will also be attenuated. The 1.5-$\mu$s duration time of control pulses is optimized by the spin-wave AFC efficiency. One primary reason for the low spin transfer efficiency is the low oscillator strength of the chosen transition. Nevertheless, our result indicates the advantages of employing the waveguide structure since such experiment performed with the bulk material would require an unrealistic laser power of the control pulse (\textgreater 10 W). Waveguides with smaller guided mode (such as type-I waveguides \cite{reid2018optica,frqmulti2019}) may help improve the spin-wave storage efficiency. 

Because of the imperfect control pulses, the two-level AFC echoes are not completely suppressed, leading to a conditional on-demand storage of 100 temporal modes. There must be a time delay (the spin-wave storage time) after several input modes are continuously injected to avoid the circumstance that the remaining AFC echoes overlap with the spin-wave AFC echoes. For example, given the spin-wave storage time of 1.5 $\mu$s here, no more than 7 (the spin-wave storage time of 1.5 $\mu$s / the mode size of 200 ns) occupied modes can be actually stored. After that time delay, there is still time to inject the remaining modes in groups of no more than 7 modes. The multimode capacity of spin-wave storage is currently limited by the fast spin dephasing. Unconditional spin-wave storage of 200 modes can be obtained by employing the dynamical decoupling to overcome the spin dephasing \cite{2020DD053,ma2021onehour}. Since our on-chip waveguide can be easily combined with on-chip coplanar electrical waveguide \cite{kindem2020control}, fast and efficient spin manipulations could be expected. We also notice that the efficiency decreases with the number of modes, similar as that observed in a previous work \cite{EU100-50}. This phenomenon still remains to be studied. We further implement a single-mode storage experiment to benchmark the coherent storage capacity similar as that reported in the previous section. The interference visibility after spin-wave AFC storage is $97\% \pm 3\%$ [Fig. \ref{spin} (c)], further reveals that the coherent phase is preserved in the spin-wave AFC storage process.
\section{discussion and conclusion }
An on-chip waveguide with a low insertion loss of 0.2 dB is fabricated on a $\mathrm {^{151}Eu^{3+}}$:$\mathrm{Y_2SiO_5}$ crystal using FLM. We extend the storage bandwidth of the $\mathrm {^{151}Eu^{3+}}$ ions to 11 MHz by applying a novel pumping scheme. Based on this, we demonstrate the 200-mode storage using AFC scheme and conditional 100-mode on-demand storage using the spin-wave AFC scheme. The coherent phases of the stored modes are well preserved during the storage processes. Longer storage times could be obtained by combing with on-chip electrical waveguides to accomplish dynamical decoupling to protect the spin coherence. The storage efficiency is currently limited by the low effective absorption of the sample and the low spin-wave transfer efficiency. Higher AFC efficiencies could be obtained with optimized comb structure and an impedance-matched cavity \cite{gisin2014njp}. Higher spin-wave transfer efficiencies could be obtained by fabricating a waveguide with a smaller guided mode and introducing dynamical decoupling to overcome the spin dephasing. Even without spin-wave storage, the two-level AFC storage with a large multimode capacity can still be useful in non-hierarchical quantum repeaters \cite{tit2014prl}. The multimode integrated memory demonstrated here could be an important ingredient for the construction of quantum repeater \cite{gisin2011rmp,multimodecomm,liu2021experimental,lago2021nature} and transportable quantum memory \cite{nature_6hour,ma2021onehour}, towards the ultimate goal of global-scale quantum communication.
\section{acknowledgments}

This work is supported by the National Key R\&D Program of China (No. 2017YFA0304100), Innovation Program for Quantum Science and Technology (No. 2021ZD0301200), the National Natural Science Foundation of China (Nos. 11774331, 11774335, 11821404 and 11654002) and the Fundamental Research Funds for the Central Universities (No. WK2470000026 and No. WK2470000029). Z.-Q.Z acknowledges the support from the Youth Innovation Promotion Association CAS.
\appendix

\section{Waveguide coupling and losses}

The method of waveguide coupling is as follows. The output beam from the FC has a diameter of 2 mm. 3 lenses modify the incoming and outcoupled beams as follows. The first two lenses have a focal length of 100 mm and 150 mm respectively and a distance of 250 mm, forming a beam expander, which expands the beam diameter from 2 mm to 3 mm. The third 75-mm lens focuses the beam into the waveguide with the help of a 3-axis translator. The focused beam has a diameter of approximately 10.9 $\mu$m. By carefully adjusting the 75-mm lens, the diameter of the focused beam is modified to match the mode of the waveguide. After the waveguide, there are 75-mm, 150-mm and 100-mm lenses in sequence to couple the laser into the FC.

The measurement of the insertion loss of the waveguide is implemented as follows. First, we carefully couple the laser into our waveguide, similar to that mentioned above but without the cryostat. To ensure the measured beam is actually guided in the waveguide, we check it in two ways. On the one hand, at the exit side of the waveguide, an aperture is placed between the 150-mm lens and 100-mm lens. The aperture is gradually closed until the single-mode coupling efficiency begins to 
decrease. On the other hand, a reflector is placed behind the outcoupled 100-mm lens to reflect the output beam of the waveguide onto a light plate. Looking at the pattern on the light plate, and adjusting the aperture so that only the guided mode can pass through. In our case, the power passing through the aperture is measured as 1.15 mW. After that, we remove the aperture and move the crystal to couple the laser into the bulk regime. The measured power is 1.21 mW after transmission. Finally, the insertion loss of the waveguide is calculated as ${IL} = -10\times\log\frac{1.15}{1.21} = 0.22$ dB. The transmission loss is 0.15 dB/cm.

The transmission of the cryostat alone is 94.9\%. The transmission of the bulk regime of the crystal is 82.7\%, which is consistent with the calculted Fresnel loss ($\sim$ 16\% for YSO crystal under normal incidence). We note that, here we assume that the Fresnel loss is the same for the waveguide and the bulk regime. However, the modified refractive index of the waveguide regime may lead to a different Fresnel loss in practice. As a result, the transmission loss may be underestimated here. The total transmission through the cryostat and the waveguide (bulk) regime of the crystal is 72.6\% (77.6\%). Therefore, the transmission of the waveguide alone is inferred to be 94\%, i.e. insertion losses of 0.2 dB, which is consistent with the independently measured IL mentioned above.\begin{figure}[htbp]
\centering
\includegraphics[width=0.85\linewidth]{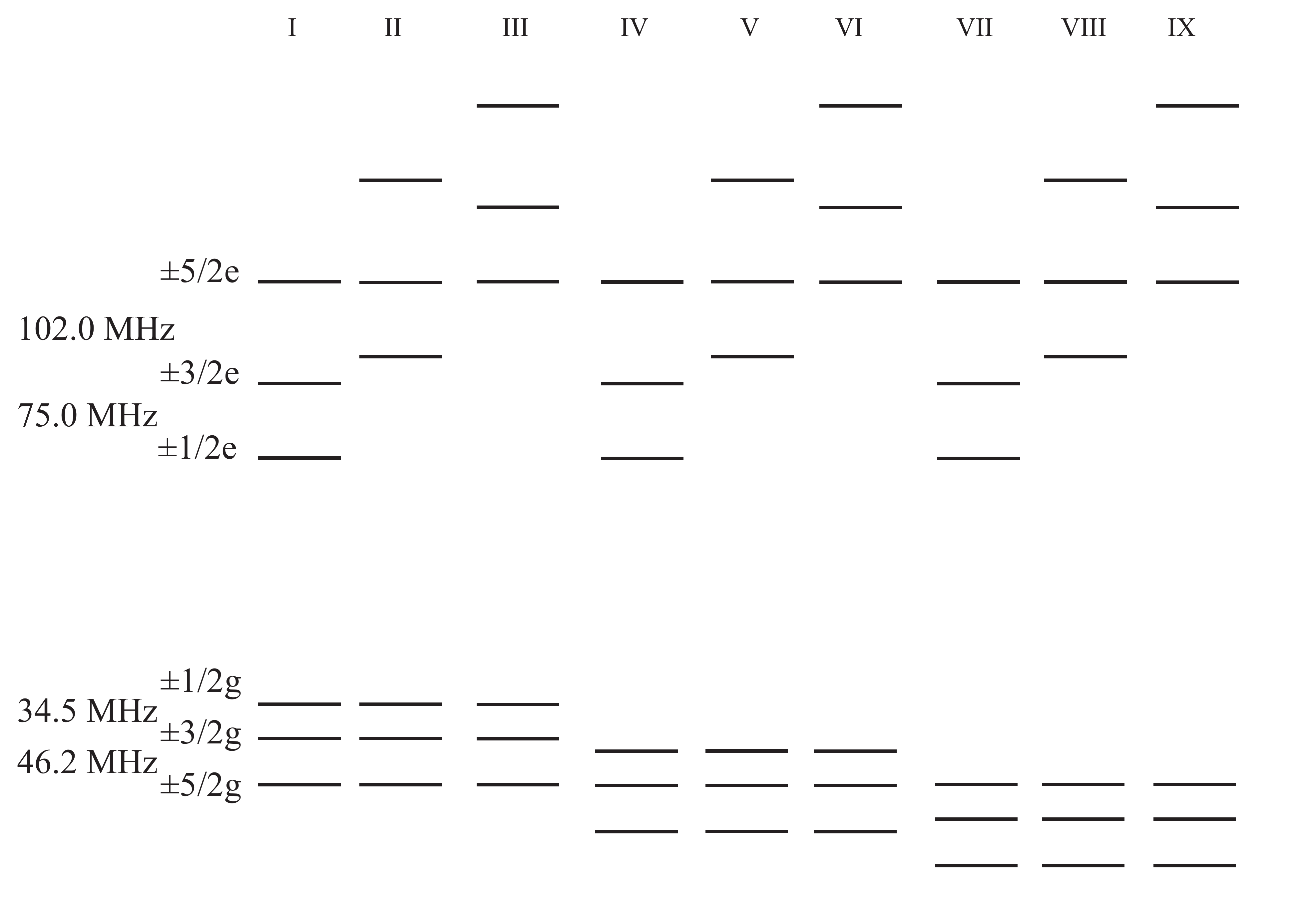}
\caption{The aligned energy diagrams of each class of ions.}
\label{appa}
\end{figure} The transmission of the lenses that between the output window of the cryostat and the fibre coupler is 93.2\%. The fibre coupling efficiency is 82.7\%. The overall transmission efficiency from the front of the cryostat to the single-mode fibre is 56\%. 

\section{The energy diagrams of nine classes of ions of $\mathrm {^{151}Eu^{3+}}$ in $\mathrm {Y_2SiO_5}$ crystal.}

The aligned energy diagrams (Fig. \ref{appa}) could help to understand the stick diagram in Fig. \ref{levelpump} (b). For example, setting the frequency of transition ${\left | \pm {1/2} \right \rangle}_g\rightarrow{\left | \pm {1/2} \right \rangle}_e$ of class-I ions as zero, the relative frequency of the transition ${\left | \pm {3/2} \right \rangle}_g\rightarrow{\left | \pm {5/2} \right \rangle}_e$ of class-VIII ions is 34.5 + 46.2 + 34.5 + 102 + 75 + 102 = 394.2 (MHz).


\begin{thebibliography}{43}%
\makeatletter
\providecommand \@ifxundefined [1]{%
 \@ifx{#1\undefined}
}%
\providecommand \@ifnum [1]{%
 \ifnum #1\expandafter \@firstoftwo
 \else \expandafter \@secondoftwo
 \fi
}%
\providecommand \@ifx [1]{%
 \ifx #1\expandafter \@firstoftwo
 \else \expandafter \@secondoftwo
 \fi
}%
\providecommand \natexlab [1]{#1}%
\providecommand \enquote  [1]{``#1''}%
\providecommand \bibnamefont  [1]{#1}%
\providecommand \bibfnamefont [1]{#1}%
\providecommand \citenamefont [1]{#1}%
\providecommand \href@noop [0]{\@secondoftwo}%
\providecommand \href [0]{\begingroup \@sanitize@url \@href}%
\providecommand \@href[1]{\@@startlink{#1}\@@href}%
\providecommand \@@href[1]{\endgroup#1\@@endlink}%
\providecommand \@sanitize@url [0]{\catcode `\\12\catcode `\$12\catcode
  `\&12\catcode `\#12\catcode `\^12\catcode `\_12\catcode `\%12\relax}%
\providecommand \@@startlink[1]{}%
\providecommand \@@endlink[0]{}%
\providecommand \url  [0]{\begingroup\@sanitize@url \@url }%
\providecommand \@url [1]{\endgroup\@href {#1}{\urlprefix }}%
\providecommand \urlprefix  [0]{URL }%
\providecommand \Eprint [0]{\href }%
\providecommand \doibase [0]{https://doi.org/}%
\providecommand \selectlanguage [0]{\@gobble}%
\providecommand \bibinfo  [0]{\@secondoftwo}%
\providecommand \bibfield  [0]{\@secondoftwo}%
\providecommand \translation [1]{[#1]}%
\providecommand \BibitemOpen [0]{}%
\providecommand \bibitemStop [0]{}%
\providecommand \bibitemNoStop [0]{.\EOS\space}%
\providecommand \EOS [0]{\spacefactor3000\relax}%
\providecommand \BibitemShut  [1]{\csname bibitem#1\endcsname}%
\let\auto@bib@innerbib\@empty
\bibitem [{\citenamefont {de~Riedmatten}\ \emph {et~al.}(2008)\citenamefont
  {de~Riedmatten}, \citenamefont {Afzelius}, \citenamefont {Staudt},
  \citenamefont {Simon},\ and\ \citenamefont {Gisin}}]{gisin2008nature}%
  \BibitemOpen
  \bibfield  {author} {\bibinfo {author} {\bibfnamefont {H.}~\bibnamefont
  {de~Riedmatten}}, \bibinfo {author} {\bibfnamefont {M.}~\bibnamefont
  {Afzelius}}, \bibinfo {author} {\bibfnamefont {M.~U.}\ \bibnamefont
  {Staudt}}, \bibinfo {author} {\bibfnamefont {C.}~\bibnamefont {Simon}},\ and\
  \bibinfo {author} {\bibfnamefont {N.}~\bibnamefont {Gisin}},\ }\href {<Go to
  ISI>://WOS:000261559900045 https://www.nature.com/articles/nature07607.pdf}
  {\bibfield  {journal} {\bibinfo  {journal} {Nature}\ }\textbf {\bibinfo
  {volume} {456}},\ \bibinfo {pages} {773} (\bibinfo {year}
  {2008})}\BibitemShut {NoStop}%
\bibitem [{\citenamefont {Sangouard}\ \emph {et~al.}(2011)\citenamefont
  {Sangouard}, \citenamefont {Simon}, \citenamefont {de~Riedmatten},\ and\
  \citenamefont {Gisin}}]{gisin2011rmp}%
  \BibitemOpen
  \bibfield  {author} {\bibinfo {author} {\bibfnamefont {N.}~\bibnamefont
  {Sangouard}}, \bibinfo {author} {\bibfnamefont {C.}~\bibnamefont {Simon}},
  \bibinfo {author} {\bibfnamefont {H.}~\bibnamefont {de~Riedmatten}},\ and\
  \bibinfo {author} {\bibfnamefont {N.}~\bibnamefont {Gisin}},\ }\href
  {https://doi.org/10.1103/RevModPhys.83.33} {\bibfield  {journal} {\bibinfo
  {journal} {Rev. Mod. Phys.}\ }\textbf {\bibinfo {volume} {83}},\ \bibinfo
  {pages} {33} (\bibinfo {year} {2011})}\BibitemShut {NoStop}%
\bibitem [{\citenamefont {B\"ottger}\ \emph {et~al.}(2009)\citenamefont
  {B\"ottger}, \citenamefont {Thiel}, \citenamefont {Cone},\ and\ \citenamefont
  {Sun}}]{Er2009prb}%
  \BibitemOpen
  \bibfield  {author} {\bibinfo {author} {\bibfnamefont {T.}~\bibnamefont
  {B\"ottger}}, \bibinfo {author} {\bibfnamefont {C.~W.}\ \bibnamefont
  {Thiel}}, \bibinfo {author} {\bibfnamefont {R.~L.}\ \bibnamefont {Cone}},\
  and\ \bibinfo {author} {\bibfnamefont {Y.}~\bibnamefont {Sun}},\ }\href
  {https://doi.org/10.1103/PhysRevB.79.115104} {\bibfield  {journal} {\bibinfo
  {journal} {Phys. Rev. B}\ }\textbf {\bibinfo {volume} {79}},\ \bibinfo
  {pages} {115104} (\bibinfo {year} {2009})}\BibitemShut {NoStop}%
\bibitem [{\citenamefont {Zhong}\ \emph {et~al.}(2015)\citenamefont {Zhong},
  \citenamefont {Hedges}, \citenamefont {Ahlefeldt}, \citenamefont
  {Bartholomew}, \citenamefont {Beavan}, \citenamefont {Wittig}, \citenamefont
  {Longdell},\ and\ \citenamefont {Sellars}}]{nature_6hour}%
  \BibitemOpen
  \bibfield  {author} {\bibinfo {author} {\bibfnamefont {M.}~\bibnamefont
  {Zhong}}, \bibinfo {author} {\bibfnamefont {M.~P.}\ \bibnamefont {Hedges}},
  \bibinfo {author} {\bibfnamefont {R.~L.}\ \bibnamefont {Ahlefeldt}}, \bibinfo
  {author} {\bibfnamefont {J.~G.}\ \bibnamefont {Bartholomew}}, \bibinfo
  {author} {\bibfnamefont {S.~E.}\ \bibnamefont {Beavan}}, \bibinfo {author}
  {\bibfnamefont {S.~M.}\ \bibnamefont {Wittig}}, \bibinfo {author}
  {\bibfnamefont {J.~J.}\ \bibnamefont {Longdell}},\ and\ \bibinfo {author}
  {\bibfnamefont {M.~J.}\ \bibnamefont {Sellars}},\ }\href
  {https://doi.org/10.1038/nature14025} {\bibfield  {journal} {\bibinfo
  {journal} {Nature}\ }\textbf {\bibinfo {volume} {517}},\ \bibinfo {pages}
  {177} (\bibinfo {year} {2015})}\BibitemShut {NoStop}%
\bibitem [{\citenamefont {Sinclair}\ \emph {et~al.}(2017)\citenamefont
  {Sinclair}, \citenamefont {Oblak}, \citenamefont {Thiel}, \citenamefont
  {Cone},\ and\ \citenamefont {Tittel}}]{tit2017prl}%
  \BibitemOpen
  \bibfield  {author} {\bibinfo {author} {\bibfnamefont {N.}~\bibnamefont
  {Sinclair}}, \bibinfo {author} {\bibfnamefont {D.}~\bibnamefont {Oblak}},
  \bibinfo {author} {\bibfnamefont {C.~W.}\ \bibnamefont {Thiel}}, \bibinfo
  {author} {\bibfnamefont {R.~L.}\ \bibnamefont {Cone}},\ and\ \bibinfo
  {author} {\bibfnamefont {W.}~\bibnamefont {Tittel}},\ }\href
  {https://doi.org/10.1103/PhysRevLett.118.100504} {\bibfield  {journal}
  {\bibinfo  {journal} {Phys. Rev. Lett.}\ }\textbf {\bibinfo {volume} {118}},\
  \bibinfo {pages} {100504} (\bibinfo {year} {2017})}\BibitemShut {NoStop}%
\bibitem [{\citenamefont {Simon}\ \emph {et~al.}(2007)\citenamefont {Simon},
  \citenamefont {de~Riedmatten}, \citenamefont {Afzelius}, \citenamefont
  {Sangouard}, \citenamefont {Zbinden},\ and\ \citenamefont
  {Gisin}}]{multimodecomm}%
  \BibitemOpen
  \bibfield  {author} {\bibinfo {author} {\bibfnamefont {C.}~\bibnamefont
  {Simon}}, \bibinfo {author} {\bibfnamefont {H.}~\bibnamefont
  {de~Riedmatten}}, \bibinfo {author} {\bibfnamefont {M.}~\bibnamefont
  {Afzelius}}, \bibinfo {author} {\bibfnamefont {N.}~\bibnamefont {Sangouard}},
  \bibinfo {author} {\bibfnamefont {H.}~\bibnamefont {Zbinden}},\ and\ \bibinfo
  {author} {\bibfnamefont {N.}~\bibnamefont {Gisin}},\ }\href
  {https://doi.org/10.1103/PhysRevLett.98.190503} {\bibfield  {journal}
  {\bibinfo  {journal} {Phys. Rev. Lett.}\ }\textbf {\bibinfo {volume} {98}},\
  \bibinfo {pages} {190503} (\bibinfo {year} {2007})}\BibitemShut {NoStop}%
\bibitem [{\citenamefont {Lago-Rivera}\ \emph {et~al.}(2021)\citenamefont
  {Lago-Rivera}, \citenamefont {Grandi}, \citenamefont {Rakonjac},
  \citenamefont {Seri},\ and\ \citenamefont {de~Riedmatten}}]{lago2021nature}%
  \BibitemOpen
  \bibfield  {author} {\bibinfo {author} {\bibfnamefont {D.}~\bibnamefont
  {Lago-Rivera}}, \bibinfo {author} {\bibfnamefont {S.}~\bibnamefont {Grandi}},
  \bibinfo {author} {\bibfnamefont {J.~V.}\ \bibnamefont {Rakonjac}}, \bibinfo
  {author} {\bibfnamefont {A.}~\bibnamefont {Seri}},\ and\ \bibinfo {author}
  {\bibfnamefont {H.}~\bibnamefont {de~Riedmatten}},\ }\href@noop {} {\bibfield
   {journal} {\bibinfo  {journal} {Nature}\ }\textbf {\bibinfo {volume}
  {594}},\ \bibinfo {pages} {37} (\bibinfo {year} {2021})}\BibitemShut
  {NoStop}%
\bibitem [{\citenamefont {Liu}\ \emph {et~al.}(2021)\citenamefont {Liu},
  \citenamefont {Hu}, \citenamefont {Li}, \citenamefont {Li}, \citenamefont
  {Li}, \citenamefont {Liang}, \citenamefont {Zhou}, \citenamefont {Li},\ and\
  \citenamefont {Guo}}]{liu2021experimental}%
  \BibitemOpen
  \bibfield  {author} {\bibinfo {author} {\bibfnamefont {X.}~\bibnamefont
  {Liu}}, \bibinfo {author} {\bibfnamefont {J.}~\bibnamefont {Hu}}, \bibinfo
  {author} {\bibfnamefont {Z.-F.}\ \bibnamefont {Li}}, \bibinfo {author}
  {\bibfnamefont {X.}~\bibnamefont {Li}}, \bibinfo {author} {\bibfnamefont
  {P.-Y.}\ \bibnamefont {Li}}, \bibinfo {author} {\bibfnamefont {P.-J.}\
  \bibnamefont {Liang}}, \bibinfo {author} {\bibfnamefont {Z.-Q.}\ \bibnamefont
  {Zhou}}, \bibinfo {author} {\bibfnamefont {C.-F.}\ \bibnamefont {Li}},\ and\
  \bibinfo {author} {\bibfnamefont {G.-C.}\ \bibnamefont {Guo}},\ }\href@noop
  {} {\bibfield  {journal} {\bibinfo  {journal} {Nature}\ }\textbf {\bibinfo
  {volume} {594}},\ \bibinfo {pages} {41} (\bibinfo {year} {2021})}\BibitemShut
  {NoStop}%
\bibitem [{\citenamefont {Tang}\ \emph {et~al.}(2015)\citenamefont {Tang},
  \citenamefont {Zhou}, \citenamefont {Wang}, \citenamefont {Li}, \citenamefont
  {Liu}, \citenamefont {Hua}, \citenamefont {Zou}, \citenamefont {Wang},
  \citenamefont {He}, \citenamefont {Chen}, \citenamefont {Sun}, \citenamefont
  {Yu}, \citenamefont {Li}, \citenamefont {Zha}, \citenamefont {Ni},
  \citenamefont {Niu}, \citenamefont {Li},\ and\ \citenamefont
  {Guo}}]{zzq2015nc}%
  \BibitemOpen
  \bibfield  {author} {\bibinfo {author} {\bibfnamefont {J.~S.}\ \bibnamefont
  {Tang}}, \bibinfo {author} {\bibfnamefont {Z.~Q.}\ \bibnamefont {Zhou}},
  \bibinfo {author} {\bibfnamefont {Y.~T.}\ \bibnamefont {Wang}}, \bibinfo
  {author} {\bibfnamefont {Y.~L.}\ \bibnamefont {Li}}, \bibinfo {author}
  {\bibfnamefont {X.}~\bibnamefont {Liu}}, \bibinfo {author} {\bibfnamefont
  {Y.~L.}\ \bibnamefont {Hua}}, \bibinfo {author} {\bibfnamefont
  {Y.}~\bibnamefont {Zou}}, \bibinfo {author} {\bibfnamefont {S.}~\bibnamefont
  {Wang}}, \bibinfo {author} {\bibfnamefont {D.~Y.}\ \bibnamefont {He}},
  \bibinfo {author} {\bibfnamefont {G.}~\bibnamefont {Chen}}, \bibinfo {author}
  {\bibfnamefont {Y.~N.}\ \bibnamefont {Sun}}, \bibinfo {author} {\bibfnamefont
  {Y.}~\bibnamefont {Yu}}, \bibinfo {author} {\bibfnamefont {M.~F.}\
  \bibnamefont {Li}}, \bibinfo {author} {\bibfnamefont {G.~W.}\ \bibnamefont
  {Zha}}, \bibinfo {author} {\bibfnamefont {H.~Q.}\ \bibnamefont {Ni}},
  \bibinfo {author} {\bibfnamefont {Z.~C.}\ \bibnamefont {Niu}}, \bibinfo
  {author} {\bibfnamefont {C.~F.}\ \bibnamefont {Li}},\ and\ \bibinfo {author}
  {\bibfnamefont {G.~C.}\ \bibnamefont {Guo}},\ }\href
  {https://doi.org/10.1038/ncomms9652} {\bibfield  {journal} {\bibinfo
  {journal} {Nature Communications}\ }\textbf {\bibinfo {volume} {6}},\
  \bibinfo {pages} {8652} (\bibinfo {year} {2015})}\BibitemShut {NoStop}%
\bibitem [{\citenamefont {Bonarota}\ \emph {et~al.}(2011)\citenamefont
  {Bonarota}, \citenamefont {Le~Gou{\"e}t},\ and\ \citenamefont
  {Chaneliere}}]{1060}%
  \BibitemOpen
  \bibfield  {author} {\bibinfo {author} {\bibfnamefont {M.}~\bibnamefont
  {Bonarota}}, \bibinfo {author} {\bibfnamefont {J.}~\bibnamefont
  {Le~Gou{\"e}t}},\ and\ \bibinfo {author} {\bibfnamefont {T.}~\bibnamefont
  {Chaneliere}},\ }\href@noop {} {\bibfield  {journal} {\bibinfo  {journal}
  {New Journal of Physics}\ }\textbf {\bibinfo {volume} {13}},\ \bibinfo
  {pages} {013013} (\bibinfo {year} {2011})}\BibitemShut {NoStop}%
\bibitem [{\citenamefont {Businger}\ \emph {et~al.}(2020)\citenamefont
  {Businger}, \citenamefont {Tiranov}, \citenamefont {Kaczmarek}, \citenamefont
  {Welinski}, \citenamefont {Zhang}, \citenamefont {Ferrier}, \citenamefont
  {Goldner},\ and\ \citenamefont {Afzelius}}]{Ybspin}%
  \BibitemOpen
  \bibfield  {author} {\bibinfo {author} {\bibfnamefont {M.}~\bibnamefont
  {Businger}}, \bibinfo {author} {\bibfnamefont {A.}~\bibnamefont {Tiranov}},
  \bibinfo {author} {\bibfnamefont {K.~T.}\ \bibnamefont {Kaczmarek}}, \bibinfo
  {author} {\bibfnamefont {S.}~\bibnamefont {Welinski}}, \bibinfo {author}
  {\bibfnamefont {Z.}~\bibnamefont {Zhang}}, \bibinfo {author} {\bibfnamefont
  {A.}~\bibnamefont {Ferrier}}, \bibinfo {author} {\bibfnamefont
  {P.}~\bibnamefont {Goldner}},\ and\ \bibinfo {author} {\bibfnamefont
  {M.}~\bibnamefont {Afzelius}},\ }\href
  {https://doi.org/10.1103/PhysRevLett.124.053606} {\bibfield  {journal}
  {\bibinfo  {journal} {Phys. Rev. Lett.}\ }\textbf {\bibinfo {volume} {124}},\
  \bibinfo {pages} {053606} (\bibinfo {year} {2020})}\BibitemShut {NoStop}%
\bibitem [{\citenamefont {Afzelius}\ \emph
  {et~al.}(2010{\natexlab{a}})\citenamefont {Afzelius}, \citenamefont {Usmani},
  \citenamefont {Amari}, \citenamefont {Lauritzen}, \citenamefont {Walther},
  \citenamefont {Simon}, \citenamefont {Sangouard}, \citenamefont
  {Min\'a\ifmmode~\check{r}\else \v{r}\fi{}}, \citenamefont {de~Riedmatten},
  \citenamefont {Gisin},\ and\ \citenamefont {Kr\"oll}}]{spin2010}%
  \BibitemOpen
  \bibfield  {author} {\bibinfo {author} {\bibfnamefont {M.}~\bibnamefont
  {Afzelius}}, \bibinfo {author} {\bibfnamefont {I.}~\bibnamefont {Usmani}},
  \bibinfo {author} {\bibfnamefont {A.}~\bibnamefont {Amari}}, \bibinfo
  {author} {\bibfnamefont {B.}~\bibnamefont {Lauritzen}}, \bibinfo {author}
  {\bibfnamefont {A.}~\bibnamefont {Walther}}, \bibinfo {author} {\bibfnamefont
  {C.}~\bibnamefont {Simon}}, \bibinfo {author} {\bibfnamefont
  {N.}~\bibnamefont {Sangouard}}, \bibinfo {author} {\bibfnamefont {J.~c.~v.}\
  \bibnamefont {Min\'a\ifmmode~\check{r}\else \v{r}\fi{}}}, \bibinfo {author}
  {\bibfnamefont {H.}~\bibnamefont {de~Riedmatten}}, \bibinfo {author}
  {\bibfnamefont {N.}~\bibnamefont {Gisin}},\ and\ \bibinfo {author}
  {\bibfnamefont {S.}~\bibnamefont {Kr\"oll}},\ }\href
  {https://doi.org/10.1103/PhysRevLett.104.040503} {\bibfield  {journal}
  {\bibinfo  {journal} {Phys. Rev. Lett.}\ }\textbf {\bibinfo {volume} {104}},\
  \bibinfo {pages} {040503} (\bibinfo {year} {2010}{\natexlab{a}})}\BibitemShut
  {NoStop}%
\bibitem [{\citenamefont {Laplane}\ \emph {et~al.}(2015)\citenamefont
  {Laplane}, \citenamefont {Jobez}, \citenamefont {Etesse}, \citenamefont
  {Timoney}, \citenamefont {Gisin},\ and\ \citenamefont
  {Afzelius}}]{Eumuti2015Q}%
  \BibitemOpen
  \bibfield  {author} {\bibinfo {author} {\bibfnamefont {C.}~\bibnamefont
  {Laplane}}, \bibinfo {author} {\bibfnamefont {P.}~\bibnamefont {Jobez}},
  \bibinfo {author} {\bibfnamefont {J.}~\bibnamefont {Etesse}}, \bibinfo
  {author} {\bibfnamefont {N.}~\bibnamefont {Timoney}}, \bibinfo {author}
  {\bibfnamefont {N.}~\bibnamefont {Gisin}},\ and\ \bibinfo {author}
  {\bibfnamefont {M.}~\bibnamefont {Afzelius}},\ }\href
  {https://doi.org/10.1088/1367-2630/18/1/013006} {\bibfield  {journal}
  {\bibinfo  {journal} {New Journal of Physics}\ }\textbf {\bibinfo {volume}
  {18}},\ \bibinfo {pages} {013006} (\bibinfo {year} {2015})}\BibitemShut
  {NoStop}%
\bibitem [{\citenamefont {Jobez}\ \emph {et~al.}(2015)\citenamefont {Jobez},
  \citenamefont {Laplane}, \citenamefont {Timoney}, \citenamefont {Gisin},
  \citenamefont {Ferrier}, \citenamefont {Goldner},\ and\ \citenamefont
  {Afzelius}}]{Euspin2015Q}%
  \BibitemOpen
  \bibfield  {author} {\bibinfo {author} {\bibfnamefont {P.}~\bibnamefont
  {Jobez}}, \bibinfo {author} {\bibfnamefont {C.}~\bibnamefont {Laplane}},
  \bibinfo {author} {\bibfnamefont {N.}~\bibnamefont {Timoney}}, \bibinfo
  {author} {\bibfnamefont {N.}~\bibnamefont {Gisin}}, \bibinfo {author}
  {\bibfnamefont {A.}~\bibnamefont {Ferrier}}, \bibinfo {author} {\bibfnamefont
  {P.}~\bibnamefont {Goldner}},\ and\ \bibinfo {author} {\bibfnamefont
  {M.}~\bibnamefont {Afzelius}},\ }\href
  {https://doi.org/10.1103/PhysRevLett.114.230502} {\bibfield  {journal}
  {\bibinfo  {journal} {Phys. Rev. Lett.}\ }\textbf {\bibinfo {volume} {114}},\
  \bibinfo {pages} {230502} (\bibinfo {year} {2015})}\BibitemShut {NoStop}%
\bibitem [{\citenamefont {G\"undo\ifmmode~\breve{g}\else \u{g}\fi{}an}\ \emph
  {et~al.}(2015)\citenamefont {G\"undo\ifmmode~\breve{g}\else \u{g}\fi{}an},
  \citenamefont {Ledingham}, \citenamefont {Kutluer}, \citenamefont {Mazzera},\
  and\ \citenamefont {de~Riedmatten}}]{Prspin2015Q}%
  \BibitemOpen
  \bibfield  {author} {\bibinfo {author} {\bibfnamefont {M.}~\bibnamefont
  {G\"undo\ifmmode~\breve{g}\else \u{g}\fi{}an}}, \bibinfo {author}
  {\bibfnamefont {P.~M.}\ \bibnamefont {Ledingham}}, \bibinfo {author}
  {\bibfnamefont {K.}~\bibnamefont {Kutluer}}, \bibinfo {author} {\bibfnamefont
  {M.}~\bibnamefont {Mazzera}},\ and\ \bibinfo {author} {\bibfnamefont
  {H.}~\bibnamefont {de~Riedmatten}},\ }\href
  {https://doi.org/10.1103/PhysRevLett.114.230501} {\bibfield  {journal}
  {\bibinfo  {journal} {Phys. Rev. Lett.}\ }\textbf {\bibinfo {volume} {114}},\
  \bibinfo {pages} {230501} (\bibinfo {year} {2015})}\BibitemShut {NoStop}%
\bibitem [{\citenamefont {Ma}\ \emph {et~al.}(2021)\citenamefont {Ma},
  \citenamefont {Ma}, \citenamefont {Zhou}, \citenamefont {Li},\ and\
  \citenamefont {Guo}}]{ma2021onehour}%
  \BibitemOpen
  \bibfield  {author} {\bibinfo {author} {\bibfnamefont {Y.}~\bibnamefont
  {Ma}}, \bibinfo {author} {\bibfnamefont {Y.-Z.}\ \bibnamefont {Ma}}, \bibinfo
  {author} {\bibfnamefont {Z.-Q.}\ \bibnamefont {Zhou}}, \bibinfo {author}
  {\bibfnamefont {C.-F.}\ \bibnamefont {Li}},\ and\ \bibinfo {author}
  {\bibfnamefont {G.-C.}\ \bibnamefont {Guo}},\ }\href
  {https://doi.org/10.1038/s41467-021-22706-y} {\bibfield  {journal} {\bibinfo
  {journal} {Nature Communications}\ }\textbf {\bibinfo {volume} {12}},\
  \bibinfo {pages} {2381} (\bibinfo {year} {2021})}\BibitemShut {NoStop}%
\bibitem [{\citenamefont {Jobez}\ \emph {et~al.}(2016)\citenamefont {Jobez},
  \citenamefont {Timoney}, \citenamefont {Laplane}, \citenamefont {Etesse},
  \citenamefont {Ferrier}, \citenamefont {Goldner}, \citenamefont {Gisin},\
  and\ \citenamefont {Afzelius}}]{EU100-50}%
  \BibitemOpen
  \bibfield  {author} {\bibinfo {author} {\bibfnamefont {P.}~\bibnamefont
  {Jobez}}, \bibinfo {author} {\bibfnamefont {N.}~\bibnamefont {Timoney}},
  \bibinfo {author} {\bibfnamefont {C.}~\bibnamefont {Laplane}}, \bibinfo
  {author} {\bibfnamefont {J.}~\bibnamefont {Etesse}}, \bibinfo {author}
  {\bibfnamefont {A.}~\bibnamefont {Ferrier}}, \bibinfo {author} {\bibfnamefont
  {P.}~\bibnamefont {Goldner}}, \bibinfo {author} {\bibfnamefont
  {N.}~\bibnamefont {Gisin}},\ and\ \bibinfo {author} {\bibfnamefont
  {M.}~\bibnamefont {Afzelius}},\ }\href
  {https://doi.org/10.1103/PhysRevA.93.032327} {\bibfield  {journal} {\bibinfo
  {journal} {Phys. Rev. A}\ }\textbf {\bibinfo {volume} {93}},\ \bibinfo
  {pages} {032327} (\bibinfo {year} {2016})}\BibitemShut {NoStop}%
\bibitem [{\citenamefont {Corrielli}\ \emph {et~al.}(2016)\citenamefont
  {Corrielli}, \citenamefont {Seri}, \citenamefont {Mazzera}, \citenamefont
  {Osellame},\ and\ \citenamefont {de~Riedmatten}}]{lminter}%
  \BibitemOpen
  \bibfield  {author} {\bibinfo {author} {\bibfnamefont {G.}~\bibnamefont
  {Corrielli}}, \bibinfo {author} {\bibfnamefont {A.}~\bibnamefont {Seri}},
  \bibinfo {author} {\bibfnamefont {M.}~\bibnamefont {Mazzera}}, \bibinfo
  {author} {\bibfnamefont {R.}~\bibnamefont {Osellame}},\ and\ \bibinfo
  {author} {\bibfnamefont {H.}~\bibnamefont {de~Riedmatten}},\ }\href
  {https://doi.org/10.1103/PhysRevApplied.5.054013} {\bibfield  {journal}
  {\bibinfo  {journal} {Phys. Rev. Applied}\ }\textbf {\bibinfo {volume} {5}},\
  \bibinfo {pages} {054013} (\bibinfo {year} {2016})}\BibitemShut {NoStop}%
\bibitem [{\citenamefont {Seri}\ \emph {et~al.}(2018)\citenamefont {Seri},
  \citenamefont {Corrielli}, \citenamefont {Lago-Rivera}, \citenamefont
  {Lenhard}, \citenamefont {de~Riedmatten}, \citenamefont {Osellame},\ and\
  \citenamefont {Mazzera}}]{reid2018optica}%
  \BibitemOpen
  \bibfield  {author} {\bibinfo {author} {\bibfnamefont {A.}~\bibnamefont
  {Seri}}, \bibinfo {author} {\bibfnamefont {G.}~\bibnamefont {Corrielli}},
  \bibinfo {author} {\bibfnamefont {D.}~\bibnamefont {Lago-Rivera}}, \bibinfo
  {author} {\bibfnamefont {A.}~\bibnamefont {Lenhard}}, \bibinfo {author}
  {\bibfnamefont {H.}~\bibnamefont {de~Riedmatten}}, \bibinfo {author}
  {\bibfnamefont {R.}~\bibnamefont {Osellame}},\ and\ \bibinfo {author}
  {\bibfnamefont {M.}~\bibnamefont {Mazzera}},\ }\href
  {https://doi.org/10.1364/Optica.5.000934} {\bibfield  {journal} {\bibinfo
  {journal} {Optica}\ }\textbf {\bibinfo {volume} {5}},\ \bibinfo {pages} {934}
  (\bibinfo {year} {2018})}\BibitemShut {NoStop}%
\bibitem [{\citenamefont {Seri}\ \emph {et~al.}(2019)\citenamefont {Seri},
  \citenamefont {Lago-Rivera}, \citenamefont {Lenhard}, \citenamefont
  {Corrielli}, \citenamefont {Osellame}, \citenamefont {Mazzera},\ and\
  \citenamefont {de~Riedmatten}}]{frqmulti2019}%
  \BibitemOpen
  \bibfield  {author} {\bibinfo {author} {\bibfnamefont {A.}~\bibnamefont
  {Seri}}, \bibinfo {author} {\bibfnamefont {D.}~\bibnamefont {Lago-Rivera}},
  \bibinfo {author} {\bibfnamefont {A.}~\bibnamefont {Lenhard}}, \bibinfo
  {author} {\bibfnamefont {G.}~\bibnamefont {Corrielli}}, \bibinfo {author}
  {\bibfnamefont {R.}~\bibnamefont {Osellame}}, \bibinfo {author}
  {\bibfnamefont {M.}~\bibnamefont {Mazzera}},\ and\ \bibinfo {author}
  {\bibfnamefont {H.}~\bibnamefont {de~Riedmatten}},\ }\href
  {https://doi.org/10.1103/PhysRevLett.123.080502} {\bibfield  {journal}
  {\bibinfo  {journal} {Phys. Rev. Lett.}\ }\textbf {\bibinfo {volume} {123}},\
  \bibinfo {pages} {080502} (\bibinfo {year} {2019})}\BibitemShut {NoStop}%
\bibitem [{\citenamefont {Liu}\ \emph {et~al.}(2020{\natexlab{a}})\citenamefont
  {Liu}, \citenamefont {Zhou}, \citenamefont {Zhu}, \citenamefont {Zheng},
  \citenamefont {Jin}, \citenamefont {Liu}, \citenamefont {Li}, \citenamefont
  {Huang}, \citenamefont {Ma}, \citenamefont {Tu} \emph
  {et~al.}}]{lc2020optica}%
  \BibitemOpen
  \bibfield  {author} {\bibinfo {author} {\bibfnamefont {C.}~\bibnamefont
  {Liu}}, \bibinfo {author} {\bibfnamefont {Z.-Q.}\ \bibnamefont {Zhou}},
  \bibinfo {author} {\bibfnamefont {T.-X.}\ \bibnamefont {Zhu}}, \bibinfo
  {author} {\bibfnamefont {L.}~\bibnamefont {Zheng}}, \bibinfo {author}
  {\bibfnamefont {M.}~\bibnamefont {Jin}}, \bibinfo {author} {\bibfnamefont
  {X.}~\bibnamefont {Liu}}, \bibinfo {author} {\bibfnamefont {P.-Y.}\
  \bibnamefont {Li}}, \bibinfo {author} {\bibfnamefont {J.-Y.}\ \bibnamefont
  {Huang}}, \bibinfo {author} {\bibfnamefont {Y.}~\bibnamefont {Ma}}, \bibinfo
  {author} {\bibfnamefont {T.}~\bibnamefont {Tu}}, \emph {et~al.},\ }\href@noop
  {} {\bibfield  {journal} {\bibinfo  {journal} {Optica}\ }\textbf {\bibinfo
  {volume} {7}},\ \bibinfo {pages} {192} (\bibinfo {year}
  {2020}{\natexlab{a}})}\BibitemShut {NoStop}%
\bibitem [{\citenamefont {Zhu}\ \emph {et~al.}(2020)\citenamefont {Zhu},
  \citenamefont {Liu}, \citenamefont {Zheng}, \citenamefont {Zhou},
  \citenamefont {Li},\ and\ \citenamefont {Guo}}]{ztxpra}%
  \BibitemOpen
  \bibfield  {author} {\bibinfo {author} {\bibfnamefont {T.-X.}\ \bibnamefont
  {Zhu}}, \bibinfo {author} {\bibfnamefont {C.}~\bibnamefont {Liu}}, \bibinfo
  {author} {\bibfnamefont {L.}~\bibnamefont {Zheng}}, \bibinfo {author}
  {\bibfnamefont {Z.-Q.}\ \bibnamefont {Zhou}}, \bibinfo {author}
  {\bibfnamefont {C.-F.}\ \bibnamefont {Li}},\ and\ \bibinfo {author}
  {\bibfnamefont {G.-C.}\ \bibnamefont {Guo}},\ }\href
  {https://doi.org/10.1103/PhysRevApplied.14.054071} {\bibfield  {journal}
  {\bibinfo  {journal} {Phys. Rev. Applied}\ }\textbf {\bibinfo {volume}
  {14}},\ \bibinfo {pages} {054071} (\bibinfo {year} {2020})}\BibitemShut
  {NoStop}%
\bibitem [{\citenamefont {Liu}\ \emph {et~al.}(2020{\natexlab{b}})\citenamefont
  {Liu}, \citenamefont {Zhu}, \citenamefont {Su}, \citenamefont {Ma},
  \citenamefont {Zhou}, \citenamefont {Li},\ and\ \citenamefont {Guo}}]{lcprl}%
  \BibitemOpen
  \bibfield  {author} {\bibinfo {author} {\bibfnamefont {C.}~\bibnamefont
  {Liu}}, \bibinfo {author} {\bibfnamefont {T.-X.}\ \bibnamefont {Zhu}},
  \bibinfo {author} {\bibfnamefont {M.-X.}\ \bibnamefont {Su}}, \bibinfo
  {author} {\bibfnamefont {Y.-Z.}\ \bibnamefont {Ma}}, \bibinfo {author}
  {\bibfnamefont {Z.-Q.}\ \bibnamefont {Zhou}}, \bibinfo {author}
  {\bibfnamefont {C.-F.}\ \bibnamefont {Li}},\ and\ \bibinfo {author}
  {\bibfnamefont {G.-C.}\ \bibnamefont {Guo}},\ }\href
  {https://doi.org/10.1103/PhysRevLett.125.260504} {\bibfield  {journal}
  {\bibinfo  {journal} {Phys. Rev. Lett.}\ }\textbf {\bibinfo {volume} {125}},\
  \bibinfo {pages} {260504} (\bibinfo {year} {2020}{\natexlab{b}})}\BibitemShut
  {NoStop}%
\bibitem [{\citenamefont {Zhong}\ \emph {et~al.}(2017)\citenamefont {Zhong},
  \citenamefont {Kindem}, \citenamefont {Bartholomew}, \citenamefont {Rochman},
  \citenamefont {Craiciu}, \citenamefont {Miyazono}, \citenamefont
  {Bettinelli}, \citenamefont {Cavalli}, \citenamefont {Verma}, \citenamefont
  {Nam}, \citenamefont {Marsili}, \citenamefont {Shaw}, \citenamefont {Beyer},\
  and\ \citenamefont {Faraon}}]{fara2017sci}%
  \BibitemOpen
  \bibfield  {author} {\bibinfo {author} {\bibfnamefont {T.}~\bibnamefont
  {Zhong}}, \bibinfo {author} {\bibfnamefont {J.~M.}\ \bibnamefont {Kindem}},
  \bibinfo {author} {\bibfnamefont {J.~G.}\ \bibnamefont {Bartholomew}},
  \bibinfo {author} {\bibfnamefont {J.}~\bibnamefont {Rochman}}, \bibinfo
  {author} {\bibfnamefont {I.}~\bibnamefont {Craiciu}}, \bibinfo {author}
  {\bibfnamefont {E.}~\bibnamefont {Miyazono}}, \bibinfo {author}
  {\bibfnamefont {M.}~\bibnamefont {Bettinelli}}, \bibinfo {author}
  {\bibfnamefont {E.}~\bibnamefont {Cavalli}}, \bibinfo {author} {\bibfnamefont
  {V.}~\bibnamefont {Verma}}, \bibinfo {author} {\bibfnamefont {S.~W.}\
  \bibnamefont {Nam}}, \bibinfo {author} {\bibfnamefont {F.}~\bibnamefont
  {Marsili}}, \bibinfo {author} {\bibfnamefont {M.~D.}\ \bibnamefont {Shaw}},
  \bibinfo {author} {\bibfnamefont {A.~D.}\ \bibnamefont {Beyer}},\ and\
  \bibinfo {author} {\bibfnamefont {A.}~\bibnamefont {Faraon}},\ }\href
  {https://doi.org/10.1126/science.aan5959} {\bibfield  {journal} {\bibinfo
  {journal} {Science}\ }\textbf {\bibinfo {volume} {357}},\ \bibinfo {pages}
  {1392} (\bibinfo {year} {2017})}\BibitemShut {NoStop}%
\bibitem [{\citenamefont {Craiciu}\ \emph {et~al.}(2019)\citenamefont
  {Craiciu}, \citenamefont {Lei}, \citenamefont {Rochman}, \citenamefont
  {Kindem}, \citenamefont {Bartholomew}, \citenamefont {Miyazono},
  \citenamefont {Zhong}, \citenamefont {Sinclair},\ and\ \citenamefont
  {Faraon}}]{fara2019prap}%
  \BibitemOpen
  \bibfield  {author} {\bibinfo {author} {\bibfnamefont {I.}~\bibnamefont
  {Craiciu}}, \bibinfo {author} {\bibfnamefont {M.}~\bibnamefont {Lei}},
  \bibinfo {author} {\bibfnamefont {J.}~\bibnamefont {Rochman}}, \bibinfo
  {author} {\bibfnamefont {J.~M.}\ \bibnamefont {Kindem}}, \bibinfo {author}
  {\bibfnamefont {J.~G.}\ \bibnamefont {Bartholomew}}, \bibinfo {author}
  {\bibfnamefont {E.}~\bibnamefont {Miyazono}}, \bibinfo {author}
  {\bibfnamefont {T.}~\bibnamefont {Zhong}}, \bibinfo {author} {\bibfnamefont
  {N.}~\bibnamefont {Sinclair}},\ and\ \bibinfo {author} {\bibfnamefont
  {A.}~\bibnamefont {Faraon}},\ }\href
  {https://doi.org/10.1103/PhysRevApplied.12.024062} {\bibfield  {journal}
  {\bibinfo  {journal} {Phys. Rev. Applied}\ }\textbf {\bibinfo {volume}
  {12}},\ \bibinfo {pages} {024062} (\bibinfo {year} {2019})}\BibitemShut
  {NoStop}%
\bibitem [{\citenamefont {Kindem}\ \emph {et~al.}(2020)\citenamefont {Kindem},
  \citenamefont {Ruskuc}, \citenamefont {Bartholomew}, \citenamefont {Rochman},
  \citenamefont {Huan},\ and\ \citenamefont {Faraon}}]{kindem2020control}%
  \BibitemOpen
  \bibfield  {author} {\bibinfo {author} {\bibfnamefont {J.~M.}\ \bibnamefont
  {Kindem}}, \bibinfo {author} {\bibfnamefont {A.}~\bibnamefont {Ruskuc}},
  \bibinfo {author} {\bibfnamefont {J.~G.}\ \bibnamefont {Bartholomew}},
  \bibinfo {author} {\bibfnamefont {J.}~\bibnamefont {Rochman}}, \bibinfo
  {author} {\bibfnamefont {Y.~Q.}\ \bibnamefont {Huan}},\ and\ \bibinfo
  {author} {\bibfnamefont {A.}~\bibnamefont {Faraon}},\ }\href@noop {}
  {\bibfield  {journal} {\bibinfo  {journal} {Nature}\ }\textbf {\bibinfo
  {volume} {580}},\ \bibinfo {pages} {201} (\bibinfo {year}
  {2020})}\BibitemShut {NoStop}%
\bibitem [{\citenamefont {Staudt}\ \emph
  {et~al.}(2007{\natexlab{a}})\citenamefont {Staudt}, \citenamefont {Afzelius},
  \citenamefont {de~Riedmatten}, \citenamefont {Hastings-Simon}, \citenamefont
  {Simon}, \citenamefont {Ricken}, \citenamefont {Suche}, \citenamefont
  {Sohler},\ and\ \citenamefont {Gisin}}]{gisin2007prl}%
  \BibitemOpen
  \bibfield  {author} {\bibinfo {author} {\bibfnamefont {M.~U.}\ \bibnamefont
  {Staudt}}, \bibinfo {author} {\bibfnamefont {M.}~\bibnamefont {Afzelius}},
  \bibinfo {author} {\bibfnamefont {H.}~\bibnamefont {de~Riedmatten}}, \bibinfo
  {author} {\bibfnamefont {S.~R.}\ \bibnamefont {Hastings-Simon}}, \bibinfo
  {author} {\bibfnamefont {C.}~\bibnamefont {Simon}}, \bibinfo {author}
  {\bibfnamefont {R.}~\bibnamefont {Ricken}}, \bibinfo {author} {\bibfnamefont
  {H.}~\bibnamefont {Suche}}, \bibinfo {author} {\bibfnamefont
  {W.}~\bibnamefont {Sohler}},\ and\ \bibinfo {author} {\bibfnamefont
  {N.}~\bibnamefont {Gisin}},\ }\href
  {https://doi.org/10.1103/PhysRevLett.99.173602} {\bibfield  {journal}
  {\bibinfo  {journal} {Phys. Rev. Lett.}\ }\textbf {\bibinfo {volume} {99}},\
  \bibinfo {pages} {173602} (\bibinfo {year} {2007}{\natexlab{a}})}\BibitemShut
  {NoStop}%
\bibitem [{\citenamefont {Staudt}\ \emph
  {et~al.}(2007{\natexlab{b}})\citenamefont {Staudt}, \citenamefont
  {Hastings-Simon}, \citenamefont {Nilsson}, \citenamefont {Afzelius},
  \citenamefont {Scarani}, \citenamefont {Ricken}, \citenamefont {Suche},
  \citenamefont {Sohler}, \citenamefont {Tittel},\ and\ \citenamefont
  {Gisin}}]{tit2007prl}%
  \BibitemOpen
  \bibfield  {author} {\bibinfo {author} {\bibfnamefont {M.~U.}\ \bibnamefont
  {Staudt}}, \bibinfo {author} {\bibfnamefont {S.~R.}\ \bibnamefont
  {Hastings-Simon}}, \bibinfo {author} {\bibfnamefont {M.}~\bibnamefont
  {Nilsson}}, \bibinfo {author} {\bibfnamefont {M.}~\bibnamefont {Afzelius}},
  \bibinfo {author} {\bibfnamefont {V.}~\bibnamefont {Scarani}}, \bibinfo
  {author} {\bibfnamefont {R.}~\bibnamefont {Ricken}}, \bibinfo {author}
  {\bibfnamefont {H.}~\bibnamefont {Suche}}, \bibinfo {author} {\bibfnamefont
  {W.}~\bibnamefont {Sohler}}, \bibinfo {author} {\bibfnamefont
  {W.}~\bibnamefont {Tittel}},\ and\ \bibinfo {author} {\bibfnamefont
  {N.}~\bibnamefont {Gisin}},\ }\href
  {https://doi.org/10.1103/PhysRevLett.98.113601} {\bibfield  {journal}
  {\bibinfo  {journal} {Phys. Rev. Lett.}\ }\textbf {\bibinfo {volume} {98}},\
  \bibinfo {pages} {113601} (\bibinfo {year} {2007}{\natexlab{b}})}\BibitemShut
  {NoStop}%
\bibitem [{\citenamefont {Saglamyurek}\ \emph {et~al.}(2011)\citenamefont
  {Saglamyurek}, \citenamefont {Sinclair}, \citenamefont {Jin}, \citenamefont
  {Slater}, \citenamefont {Oblak}, \citenamefont {Bussìres}, \citenamefont
  {George}, \citenamefont {Ricken}, \citenamefont {Sohler},\ and\ \citenamefont
  {Tittel}}]{tit2011nature}%
  \BibitemOpen
  \bibfield  {author} {\bibinfo {author} {\bibfnamefont {E.}~\bibnamefont
  {Saglamyurek}}, \bibinfo {author} {\bibfnamefont {N.}~\bibnamefont
  {Sinclair}}, \bibinfo {author} {\bibfnamefont {J.}~\bibnamefont {Jin}},
  \bibinfo {author} {\bibfnamefont {J.~A.}\ \bibnamefont {Slater}}, \bibinfo
  {author} {\bibfnamefont {D.}~\bibnamefont {Oblak}}, \bibinfo {author}
  {\bibfnamefont {F.}~\bibnamefont {Bussìres}}, \bibinfo {author}
  {\bibfnamefont {M.}~\bibnamefont {George}}, \bibinfo {author} {\bibfnamefont
  {R.}~\bibnamefont {Ricken}}, \bibinfo {author} {\bibfnamefont
  {W.}~\bibnamefont {Sohler}},\ and\ \bibinfo {author} {\bibfnamefont
  {W.}~\bibnamefont {Tittel}},\ }\href {https://doi.org/10.1038/nature09719}
  {\bibfield  {journal} {\bibinfo  {journal} {Nature}\ }\textbf {\bibinfo
  {volume} {469}},\ \bibinfo {pages} {512} (\bibinfo {year}
  {2011})}\BibitemShut {NoStop}%
\bibitem [{\citenamefont {Sinclair}\ \emph {et~al.}(2014)\citenamefont
  {Sinclair}, \citenamefont {Saglamyurek}, \citenamefont {Mallahzadeh},
  \citenamefont {Slater}, \citenamefont {George}, \citenamefont {Ricken},
  \citenamefont {Hedges}, \citenamefont {Oblak}, \citenamefont {Simon},
  \citenamefont {Sohler},\ and\ \citenamefont {Tittel}}]{tit2014prl}%
  \BibitemOpen
  \bibfield  {author} {\bibinfo {author} {\bibfnamefont {N.}~\bibnamefont
  {Sinclair}}, \bibinfo {author} {\bibfnamefont {E.}~\bibnamefont
  {Saglamyurek}}, \bibinfo {author} {\bibfnamefont {H.}~\bibnamefont
  {Mallahzadeh}}, \bibinfo {author} {\bibfnamefont {J.~A.}\ \bibnamefont
  {Slater}}, \bibinfo {author} {\bibfnamefont {M.}~\bibnamefont {George}},
  \bibinfo {author} {\bibfnamefont {R.}~\bibnamefont {Ricken}}, \bibinfo
  {author} {\bibfnamefont {M.~P.}\ \bibnamefont {Hedges}}, \bibinfo {author}
  {\bibfnamefont {D.}~\bibnamefont {Oblak}}, \bibinfo {author} {\bibfnamefont
  {C.}~\bibnamefont {Simon}}, \bibinfo {author} {\bibfnamefont
  {W.}~\bibnamefont {Sohler}},\ and\ \bibinfo {author} {\bibfnamefont
  {W.}~\bibnamefont {Tittel}},\ }\href
  {https://doi.org/10.1103/PhysRevLett.113.053603} {\bibfield  {journal}
  {\bibinfo  {journal} {Phys. Rev. Lett.}\ }\textbf {\bibinfo {volume} {113}},\
  \bibinfo {pages} {053603} (\bibinfo {year} {2014})}\BibitemShut {NoStop}%
\bibitem [{\citenamefont {Askarani}\ \emph {et~al.}(2019)\citenamefont
  {Askarani}, \citenamefont {Puigibert}, \citenamefont {Lutz}, \citenamefont
  {Verma}, \citenamefont {Shaw}, \citenamefont {Nam}, \citenamefont {Sinclair},
  \citenamefont {Oblak},\ and\ \citenamefont {Tittel}}]{tit2019prap}%
  \BibitemOpen
  \bibfield  {author} {\bibinfo {author} {\bibfnamefont {M.~F.}\ \bibnamefont
  {Askarani}}, \bibinfo {author} {\bibfnamefont {Marcel.liGrimau}\ \bibnamefont
  {Puigibert}}, \bibinfo {author} {\bibfnamefont {T.}~\bibnamefont {Lutz}},
  \bibinfo {author} {\bibfnamefont {V.~B.}\ \bibnamefont {Verma}}, \bibinfo
  {author} {\bibfnamefont {M.~D.}\ \bibnamefont {Shaw}}, \bibinfo {author}
  {\bibfnamefont {S.~W.}\ \bibnamefont {Nam}}, \bibinfo {author} {\bibfnamefont
  {N.}~\bibnamefont {Sinclair}}, \bibinfo {author} {\bibfnamefont
  {D.}~\bibnamefont {Oblak}},\ and\ \bibinfo {author} {\bibfnamefont
  {W.}~\bibnamefont {Tittel}},\ }\href
  {https://doi.org/10.1103/PhysRevApplied.11.054056} {\bibfield  {journal}
  {\bibinfo  {journal} {Phys. Rev. Applied}\ }\textbf {\bibinfo {volume}
  {11}},\ \bibinfo {pages} {054056} (\bibinfo {year} {2019})}\BibitemShut
  {NoStop}%
\bibitem [{\citenamefont {Saglamyurek}\ \emph {et~al.}(2016)\citenamefont
  {Saglamyurek}, \citenamefont {Puigibert}, \citenamefont {Zhou}, \citenamefont
  {Giner}, \citenamefont {Marsili}, \citenamefont {Verma}, \citenamefont {Nam},
  \citenamefont {Oesterling}, \citenamefont {Nippa}, \citenamefont {Oblak}
  \emph {et~al.}}]{fiberNC}%
  \BibitemOpen
  \bibfield  {author} {\bibinfo {author} {\bibfnamefont {E.}~\bibnamefont
  {Saglamyurek}}, \bibinfo {author} {\bibfnamefont {M.~G.}\ \bibnamefont
  {Puigibert}}, \bibinfo {author} {\bibfnamefont {Q.}~\bibnamefont {Zhou}},
  \bibinfo {author} {\bibfnamefont {L.}~\bibnamefont {Giner}}, \bibinfo
  {author} {\bibfnamefont {F.}~\bibnamefont {Marsili}}, \bibinfo {author}
  {\bibfnamefont {V.~B.}\ \bibnamefont {Verma}}, \bibinfo {author}
  {\bibfnamefont {S.~W.}\ \bibnamefont {Nam}}, \bibinfo {author} {\bibfnamefont
  {L.}~\bibnamefont {Oesterling}}, \bibinfo {author} {\bibfnamefont
  {D.}~\bibnamefont {Nippa}}, \bibinfo {author} {\bibfnamefont
  {D.}~\bibnamefont {Oblak}}, \emph {et~al.},\ }\href@noop {} {\bibfield
  {journal} {\bibinfo  {journal} {Nat. Commun.}\ }\textbf {\bibinfo {volume}
  {7}},\ \bibinfo {pages} {11202} (\bibinfo {year} {2016})}\BibitemShut
  {NoStop}%
\bibitem [{\citenamefont {Jin}\ \emph {et~al.}(2015)\citenamefont {Jin},
  \citenamefont {Saglamyurek}, \citenamefont {Puigibert}, \citenamefont
  {Verma}, \citenamefont {Marsili}, \citenamefont {Nam}, \citenamefont
  {Oblak},\ and\ \citenamefont {Tittel}}]{fiberprl}%
  \BibitemOpen
  \bibfield  {author} {\bibinfo {author} {\bibfnamefont {J.}~\bibnamefont
  {Jin}}, \bibinfo {author} {\bibfnamefont {E.}~\bibnamefont {Saglamyurek}},
  \bibinfo {author} {\bibfnamefont {Marcel.liGrimau}\ \bibnamefont {Puigibert}},
  \bibinfo {author} {\bibfnamefont {V.}~\bibnamefont {Verma}}, \bibinfo
  {author} {\bibfnamefont {F.}~\bibnamefont {Marsili}}, \bibinfo {author}
  {\bibfnamefont {S.~W.}\ \bibnamefont {Nam}}, \bibinfo {author} {\bibfnamefont
  {D.}~\bibnamefont {Oblak}},\ and\ \bibinfo {author} {\bibfnamefont
  {W.}~\bibnamefont {Tittel}},\ }\href
  {https://doi.org/10.1103/PhysRevLett.115.140501} {\bibfield  {journal}
  {\bibinfo  {journal} {Phys. Rev. Lett.}\ }\textbf {\bibinfo {volume} {115}},\
  \bibinfo {pages} {140501} (\bibinfo {year} {2015})}\BibitemShut {NoStop}%
\bibitem [{\citenamefont {Chen}\ and\ \citenamefont
  {de~Aldana}(2014)}]{CFwaveguide}%
  \BibitemOpen
  \bibfield  {author} {\bibinfo {author} {\bibfnamefont {F.}~\bibnamefont
  {Chen}}\ and\ \bibinfo {author} {\bibfnamefont {J.~R.~V.}\ \bibnamefont
  {de~Aldana}},\ }\href {https://doi.org/10.1002/lpor.201300025} {\bibfield
  {journal} {\bibinfo  {journal} {Laser and Photonics Reviews}\ }\textbf
  {\bibinfo {volume} {8}},\ \bibinfo {pages} {251} (\bibinfo {year}
  {2014})}\BibitemShut {NoStop}%
\bibitem [{\citenamefont {Lauritzen}\ \emph {et~al.}(2012)\citenamefont
  {Lauritzen}, \citenamefont {Timoney}, \citenamefont {Gisin}, \citenamefont
  {Afzelius}, \citenamefont {de~Riedmatten}, \citenamefont {Sun}, \citenamefont
  {Macfarlane},\ and\ \citenamefont {Cone}}]{Euspec}%
  \BibitemOpen
  \bibfield  {author} {\bibinfo {author} {\bibfnamefont {B.}~\bibnamefont
  {Lauritzen}}, \bibinfo {author} {\bibfnamefont {N.}~\bibnamefont {Timoney}},
  \bibinfo {author} {\bibfnamefont {N.}~\bibnamefont {Gisin}}, \bibinfo
  {author} {\bibfnamefont {M.}~\bibnamefont {Afzelius}}, \bibinfo {author}
  {\bibfnamefont {H.}~\bibnamefont {de~Riedmatten}}, \bibinfo {author}
  {\bibfnamefont {Y.}~\bibnamefont {Sun}}, \bibinfo {author} {\bibfnamefont
  {R.~M.}\ \bibnamefont {Macfarlane}},\ and\ \bibinfo {author} {\bibfnamefont
  {R.~L.}\ \bibnamefont {Cone}},\ }\href@noop {} {\bibfield  {journal}
  {\bibinfo  {journal} {Phys. Rev. B}\ }\textbf {\bibinfo {volume} {85}},\
  \bibinfo {pages} {115111} (\bibinfo {year} {2012})}\BibitemShut {NoStop}%
\bibitem [{\citenamefont {Afzelius}\ \emph {et~al.}(2009)\citenamefont
  {Afzelius}, \citenamefont {Simon}, \citenamefont {De~Riedmatten},\ and\
  \citenamefont {Gisin}}]{gisin2009pra}%
  \BibitemOpen
  \bibfield  {author} {\bibinfo {author} {\bibfnamefont {M.}~\bibnamefont
  {Afzelius}}, \bibinfo {author} {\bibfnamefont {C.}~\bibnamefont {Simon}},
  \bibinfo {author} {\bibfnamefont {H.}~\bibnamefont {de~Riedmatten}},\ and\
  \bibinfo {author} {\bibfnamefont {N.}~\bibnamefont {Gisin}},\ }\href
  {https://doi.org/10.1103/PhysRevA.79.052329} {\bibfield  {journal} {\bibinfo
  {journal} {Phys. Rev. A}\ }\textbf {\bibinfo {volume} {79}},\ \bibinfo
  {pages} {052329} (\bibinfo {year} {2009})}\BibitemShut {NoStop}%
\bibitem [{\citenamefont {Bonarota}\ \emph {et~al.}(2010)\citenamefont
  {Bonarota}, \citenamefont {Ruggiero}, \citenamefont {Gou\"et},\ and\
  \citenamefont {Chaneli\`ere}}]{PhysRevA.81.033803}%
  \BibitemOpen
  \bibfield  {author} {\bibinfo {author} {\bibfnamefont {M.}~\bibnamefont
  {Bonarota}}, \bibinfo {author} {\bibfnamefont {J.}~\bibnamefont {Ruggiero}},
  \bibinfo {author} {\bibfnamefont {J.~L.~L.}\ \bibnamefont {LeGou\"et}},\ and\
  \bibinfo {author} {\bibfnamefont {T.}~\bibnamefont {Chaneli\`ere}},\ }\href
  {https://doi.org/10.1103/PhysRevA.81.033803} {\bibfield  {journal} {\bibinfo
  {journal} {Phys. Rev. A}\ }\textbf {\bibinfo {volume} {81}},\ \bibinfo
  {pages} {033803} (\bibinfo {year} {2010})}\BibitemShut {NoStop}%
\bibitem [{\citenamefont {Timoney}\ \emph {et~al.}(2012)\citenamefont
  {Timoney}, \citenamefont {Lauritzen}, \citenamefont {Usmani}, \citenamefont
  {Afzelius},\ and\ \citenamefont {Gisin}}]{timoney2012atomic}%
  \BibitemOpen
  \bibfield  {author} {\bibinfo {author} {\bibfnamefont {N.}~\bibnamefont
  {Timoney}}, \bibinfo {author} {\bibfnamefont {B.}~\bibnamefont {Lauritzen}},
  \bibinfo {author} {\bibfnamefont {I.}~\bibnamefont {Usmani}}, \bibinfo
  {author} {\bibfnamefont {M.}~\bibnamefont {Afzelius}},\ and\ \bibinfo
  {author} {\bibfnamefont {N.}~\bibnamefont {Gisin}},\ }\href@noop {}
  {\bibfield  {journal} {\bibinfo  {journal} {Journal of Physics B: Atomic,
  Molecular and Optical Physics}\ }\textbf {\bibinfo {volume} {45}},\ \bibinfo
  {pages} {124001} (\bibinfo {year} {2012})}\BibitemShut {NoStop}%
\bibitem [{\citenamefont {G{\"u}ndo{\u{g}}an}\ \emph
  {et~al.}(2013)\citenamefont {G{\"u}ndo{\u{g}}an}, \citenamefont {Mazzera},
  \citenamefont {Ledingham}, \citenamefont {Cristiani},\ and\ \citenamefont
  {de~Riedmatten}}]{gundougan2013coherent}%
  \BibitemOpen
  \bibfield  {author} {\bibinfo {author} {\bibfnamefont {M.}~\bibnamefont
  {G{\"u}ndo{\u{g}}an}}, \bibinfo {author} {\bibfnamefont {M.}~\bibnamefont
  {Mazzera}}, \bibinfo {author} {\bibfnamefont {P.~M.}\ \bibnamefont
  {Ledingham}}, \bibinfo {author} {\bibfnamefont {M.}~\bibnamefont
  {Cristiani}},\ and\ \bibinfo {author} {\bibfnamefont {H.}~\bibnamefont
  {de~Riedmatten}},\ }\href@noop {} {\bibfield  {journal} {\bibinfo  {journal}
  {New Journal of Physics}\ }\textbf {\bibinfo {volume} {15}},\ \bibinfo
  {pages} {045012} (\bibinfo {year} {2013})}\BibitemShut {NoStop}%
\bibitem [{\citenamefont {Arcangeli}\ \emph {et~al.}(2014)\citenamefont
  {Arcangeli}, \citenamefont {Lovri\ifmmode~\acute{c}\else \'{c}\fi{}},
  \citenamefont {Tumino}, \citenamefont {Ferrier},\ and\ \citenamefont
  {Goldner}}]{21kPhysRevB.89.184305}%
  \BibitemOpen
  \bibfield  {author} {\bibinfo {author} {\bibfnamefont {A.}~\bibnamefont
  {Arcangeli}}, \bibinfo {author} {\bibfnamefont {M.}~\bibnamefont
  {Lovri\ifmmode~\acute{c}\else \'{c}\fi{}}}, \bibinfo {author} {\bibfnamefont
  {B.}~\bibnamefont {Tumino}}, \bibinfo {author} {\bibfnamefont
  {A.}~\bibnamefont {Ferrier}},\ and\ \bibinfo {author} {\bibfnamefont
  {P.}~\bibnamefont {Goldner}},\ }\href
  {https://doi.org/10.1103/PhysRevB.89.184305} {\bibfield  {journal} {\bibinfo
  {journal} {Phys. Rev. B}\ }\textbf {\bibinfo {volume} {89}},\ \bibinfo
  {pages} {184305} (\bibinfo {year} {2014})}\BibitemShut {NoStop}%
\bibitem [{\citenamefont {Lovri\ifmmode~\acute{c}\else \'{c}\fi{}}\ \emph
  {et~al.}(2011)\citenamefont {Lovri\ifmmode~\acute{c}\else \'{c}\fi{}},
  \citenamefont {Glasenapp}, \citenamefont {Suter}, \citenamefont {Tumino},
  \citenamefont {Ferrier}, \citenamefont {Goldner}, \citenamefont {Sabooni},
  \citenamefont {Rippe},\ and\ \citenamefont {Kr\"oll}}]{prPhysRevB.84.104417}%
  \BibitemOpen
  \bibfield  {author} {\bibinfo {author} {\bibfnamefont {M.}~\bibnamefont
  {Lovri\ifmmode~\acute{c}\else \'{c}\fi{}}}, \bibinfo {author} {\bibfnamefont
  {P.}~\bibnamefont {Glasenapp}}, \bibinfo {author} {\bibfnamefont
  {D.}~\bibnamefont {Suter}}, \bibinfo {author} {\bibfnamefont
  {B.}~\bibnamefont {Tumino}}, \bibinfo {author} {\bibfnamefont
  {A.}~\bibnamefont {Ferrier}}, \bibinfo {author} {\bibfnamefont
  {P.}~\bibnamefont {Goldner}}, \bibinfo {author} {\bibfnamefont
  {M.}~\bibnamefont {Sabooni}}, \bibinfo {author} {\bibfnamefont
  {L.}~\bibnamefont {Rippe}},\ and\ \bibinfo {author} {\bibfnamefont
  {S.}~\bibnamefont {Kr\"oll}},\ }\href
  {https://doi.org/10.1103/PhysRevB.84.104417} {\bibfield  {journal} {\bibinfo
  {journal} {Phys. Rev. B}\ }\textbf {\bibinfo {volume} {84}},\ \bibinfo
  {pages} {104417} (\bibinfo {year} {2011})}\BibitemShut {NoStop}%
\bibitem [{\citenamefont {Afzelius}\ \emph
  {et~al.}(2010{\natexlab{b}})\citenamefont {Afzelius}, \citenamefont {Usmani},
  \citenamefont {Amari}, \citenamefont {Lauritzen}, \citenamefont {Walther},
  \citenamefont {Simon}, \citenamefont {Sangouard}, \citenamefont
  {Min\'a\ifmmode~\check{r}\else \v{r}\fi{}}, \citenamefont {de~Riedmatten},
  \citenamefont {Gisin},\ and\ \citenamefont {Kr\"oll}}]{2010prlc}%
  \BibitemOpen
  \bibfield  {author} {\bibinfo {author} {\bibfnamefont {M.}~\bibnamefont
  {Afzelius}}, \bibinfo {author} {\bibfnamefont {I.}~\bibnamefont {Usmani}},
  \bibinfo {author} {\bibfnamefont {A.}~\bibnamefont {Amari}}, \bibinfo
  {author} {\bibfnamefont {B.}~\bibnamefont {Lauritzen}}, \bibinfo {author}
  {\bibfnamefont {A.}~\bibnamefont {Walther}}, \bibinfo {author} {\bibfnamefont
  {C.}~\bibnamefont {Simon}}, \bibinfo {author} {\bibfnamefont
  {N.}~\bibnamefont {Sangouard}}, \bibinfo {author} {\bibfnamefont {J.~c.~v.}\
  \bibnamefont {Min\'a\ifmmode~\check{r}\else \v{r}\fi{}}}, \bibinfo {author}
  {\bibfnamefont {H.}~\bibnamefont {de~Riedmatten}}, \bibinfo {author}
  {\bibfnamefont {N.}~\bibnamefont {Gisin}},\ and\ \bibinfo {author}
  {\bibfnamefont {S.}~\bibnamefont {Kr\"oll}},\ }\href
  {https://doi.org/10.1103/PhysRevLett.104.040503} {\bibfield  {journal}
  {\bibinfo  {journal} {Phys. Rev. Lett.}\ }\textbf {\bibinfo {volume} {104}},\
  \bibinfo {pages} {040503} (\bibinfo {year} {2010}{\natexlab{b}})}\BibitemShut
  {NoStop}%
\bibitem [{\citenamefont {Holz{\"a}pfel}\ \emph {et~al.}(2020)\citenamefont
  {Holz{\"a}pfel}, \citenamefont {Etesse}, \citenamefont {Kaczmarek},
  \citenamefont {Tiranov}, \citenamefont {Gisin},\ and\ \citenamefont
  {Afzelius}}]{2020DD053}%
  \BibitemOpen
  \bibfield  {author} {\bibinfo {author} {\bibfnamefont {A.}~\bibnamefont
  {Holz{\"a}pfel}}, \bibinfo {author} {\bibfnamefont {J.}~\bibnamefont
  {Etesse}}, \bibinfo {author} {\bibfnamefont {K.~T.}\ \bibnamefont
  {Kaczmarek}}, \bibinfo {author} {\bibfnamefont {A.}~\bibnamefont {Tiranov}},
  \bibinfo {author} {\bibfnamefont {N.}~\bibnamefont {Gisin}},\ and\ \bibinfo
  {author} {\bibfnamefont {M.}~\bibnamefont {Afzelius}},\ }\href@noop {}
  {\bibfield  {journal} {\bibinfo  {journal} {New Journal of Physics}\ }\textbf
  {\bibinfo {volume} {22}},\ \bibinfo {pages} {063009} (\bibinfo {year}
  {2020})}\BibitemShut 
  {NoStop}%
 \bibitem [{\citenamefont {Jobez}\ \emph {et~al.}(2014)\citenamefont {Jobez},
  \citenamefont {Usmani}, \citenamefont {Timoney}, \citenamefont {Laplane},
  \citenamefont {Gisin},\ and\ \citenamefont {Afzelius}}]{gisin2014njp}%
  \BibitemOpen
  \bibfield  {author} {\bibinfo {author} {\bibfnamefont {P.}~\bibnamefont
  {Jobez}}, \bibinfo {author} {\bibfnamefont {I.}~\bibnamefont {Usmani}},
  \bibinfo {author} {\bibfnamefont {N.}~\bibnamefont {Timoney}}, \bibinfo
  {author} {\bibfnamefont {C.}~\bibnamefont {Laplane}}, \bibinfo {author}
  {\bibfnamefont {N.}~\bibnamefont {Gisin}},\ and\ \bibinfo {author}
  {\bibfnamefont {M.}~\bibnamefont {Afzelius}},\ }\href
  {https://doi.org/10.1088/1367-2630/16/8/083005} {\bibfield  {journal}
  {\bibinfo  {journal} {New Journal of Physics}\ }\textbf {\bibinfo {volume}
  {16}},\ \bibinfo {pages} {083005} (\bibinfo {year} {2014})}\BibitemShut
  {NoStop}%
\end{thebibliography}
\end{document}